\documentclass[%
 reprint,
 amsmath,amssymb,
 aps,
 onecolumn
 pre,
]{revtex4-2}
\usepackage{graphicx}% Include figure files
\usepackage{dcolumn}% Align table columns on decimal point
\usepackage{bm}% bold math
\usepackage{lineno}
\usepackage{subfigure}
\usepackage{multirow}
\usepackage{setspace}
\usepackage{ulem}
\begin{document}

\preprint{APS/123-QED}
\title{Numerical analysis of electrohydrodynamic (EHD) instability in dielectric liquid-gas flows subjected to unipolar injection}% Force line breaks with \\%

\author{Qiang Liu}
\affiliation{%
	School of Energy Science and Engineering, Harbin Institute of Technology, Harbin 150001, PR China\\
	Key Laboratory of Aerospace Thermophysics, Ministry of Industry and Information Technology, Harbin 150001,PR China
}%

\author{Alberto T. Pérez}
\affiliation{
	Departamento de Electrónica y Electromagnetismo,Universidad de Sevilla, Sevilla, 41012 Spain% with \\
}%
\author{R. Deepak Selvakumar}
\affiliation{%
	School of Energy Science and Engineering, Harbin Institute of Technology, Harbin 150001, PR China\\
	Key Laboratory of Aerospace Thermophysics, Ministry of Industry and Information Technology, Harbin 150001,PR China
}%

\author{Pengfei Yang}
\author{Jian Wu}
\email{jian.wu@hit.edu.cn}
\affiliation{%
	School of Energy Science and Engineering, Harbin Institute of Technology, Harbin 150001, PR China\\
	Key Laboratory of Aerospace Thermophysics, Ministry of Industry and Information Technology, Harbin 150001,PR China
}%

\date{\today}% It is always \today, today,

\begin{abstract}
In this work, the electrohydrodynamic (EHD) instability induced by a unipolar charge injection is extended from a single-phase dielectric liquid to a two-phase system that consists of a liquid-air interface. 
A volume of fluid (VOF) model based two-phase solver was developed with simplified Maxwell equations implemented in the open-source platform OpenFOAM\textsuperscript{\textregistered}. The numerically obtained critical value for the linear stability matches well with the theoretical values. To highlight the effect of the slip boundary at interface, the deformation of the interface is ignored. A bifurcation diagram with hysteresis loop linking the linear and finite amplitude criteria, which is Uf = 0.059, was obtained in this situation. It is concluded that the lack of viscous effect at interface leads to a significant increase in the flow intensity, which is the reason for the smaller instability threshold in two-phase system. The presence of interface also changes the flow structure  and makes the flow vortices shift closer to the interface.
\end{abstract}

\maketitle                         %display desired

%\tableofcontents
\section{\label{sec:1}introduction}
Multi-phase electrohydrodynamics (EHD) is a complex subject that involves interactions of two or more fluids and also an external electric field. It attracts a wide range of fundamental research interest due to its complex flow structures and rich bifuractions\cite{melcher1969electrohydrodynamics,schnitzer2015taylor,Vlahovska2019,Papageorgiou2019,Dinesh2021}. This type of flow motion also plays the center role in several engineering applications, such as electrosprays, ink-jets, boiling heat transfer, and EHD pumping\cite{yudistira2010flight,mcgranaghan2014mechanisms,ganan2018review,vazquez2019depth,grassi2019new}.

In a two-phase EHD problem, stability of the interface between two fluids layers is a classical research topic started by Taylor and McEwan\cite{Taylar1965}. They gave a theoretical and experimental analysis to the instability of a perfect conducting liquid layer placed between two plane electrodes. A flourish of extending this topic from various aspects has occured in recent years. For example, it has been proved that the electric field is able to produce many different flow patterns in the absence of shear flow between layers\cite{Schaffer2000,Mondal2014,Mondal2018}. Some typical interface instabilities between two fluid layers like the Rayleigh-Taylor instability can also be controlled by electric field\cite{Cimpeanu2014,Yang2016,Yang2017}. When the shear flow is considered, the problem turns into film flow under electric field and the original instability will be affected significantly due to the enriched interplay between the competing mechanisms\cite{Ozen2006,papageorgiou2011,Dubrovina2017,Tomlin2020}. The influence of the wall topography\cite{Tseluiko2008,Tseluiko2011,Wray2013} as well as the AC/DC characteristics of the electric field\cite{Gambhire2012,Espin2013,Bandopadhyay2017} on film stability are also widely studied. 

Inspired by the experimental observation of Rose-window instability\cite{perez1996electrohydrodynamic,perez1997rose}, the gas-liquid flow subjected to charge injection has also attracted many research interests. The Rose-window instability arises when a corona discharge is applied on a liquid surface with low conductivity. The electric field pushes the charged liquid surface and a regular interface deformation like a Rose-window appears.	Intrigued by the flow pattern of Rose-window instability, Atten and Koulova-Nenova gave a linear stability analysis with a 2D model considering a layer of liquid and a layer of air between parallel plates under unipolar injection\cite{atten1996ehd,koulova1997ehd}. No criterion related to Rose-window instability was found in their work but a criterion for the EHD instability caused by bulk charge in liquid layer is obtained. The EHD instability caused by bulk charge is a wide researched topic in single-phase dielectric liquid due to its subcritical bifurcation instability\cite{nonlinear} and attractive flow pattern of the so-called charge void region\cite{chargeVoidRegion,voidRegionClosetoThecritical,convection2}. Atten and Koulova-Nenova's work showed that the deformation of the interface will decrease the critical value for the bulk charge instability compared to a single-phase dielectric fluid system. Many subsequent works tried to find the critical value corresponding to the Rose-window instability\cite{atten1998ehd,vega2002instability,chicon2006instability} based on Atten and Koulova-Nenova's work and this was finally made by Chicón and Pérez \cite{chicon2014stability}. The bulk charge instability threshold they found is close to Koulova-Nenova's result and a new criterion which is related to the Rose-window instability was obtained when the thickness of liquid layer is small enough. 

In the configuration of rose-window instability system, the interface works as a flexible electrode from the perspective of liquid layer. This flexible electrode has two key differences compared with the solid one in single-phase situation, which is the deformability and the slip boundary condition. The effect of deformability is discussed by Koulova-Nenova and Atten as mentioned before while the role of slip boundary has not been carefully studied. Based on the previous numerical studies on the single-phase EHD instability problem\cite{JFM,PRE,FVM} and the latest analytical work\cite{chicon2014stability}, this work presents a numerical analysis of EHD instability of a horizontal liquid-air interface subjected to unipolar charge injection with specifically developed finite-volume solver based on the open-source platform OpenFOAM\textsuperscript{\textregistered}. The effect of slip boundary condition on the bulk charge instability has been highlighted. The remainder of the paper is organized as follows: Description of the physical problem and the mathematical formulation is presented in the next section. The numerical methodology is described in Section~\ref{sec:3}. The results are presented and discussed in Section~\ref{sec:4}. Finally, the concluding remarks are summarized in the last section.

\section{Mathematical formulation}
\subsection{\label{sec:2A}\label{sec:2B}\label{sec:2C}Problem description and governing equations}
The flow domain shown in Fig.~\ref{fig:geometry} consists of two flat plate electrodes in parallel configuration that encloses an layer of an dielectric liquid and a air layer of thicknesses $D$ and $L$, respectively. The upper plate electrode is maintained at a higher electric potential $V$ and the bottom electrode is grounded. It is assumed that unipolar injection of positive ions takes place from the upper emitter electrode. A uniform and constant charge density $\rho_e~=\rho_{e0}$ (i.e., a homogeneous and autonomous injection) is considered at the emitter electrode.
\begin{figure}[tbh]
	\includegraphics[width=0.4\textwidth]{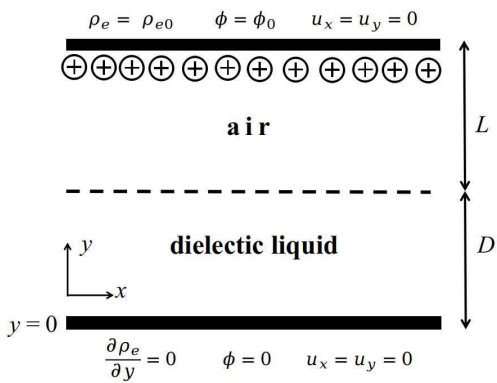}
	\caption{\label{fig:geometry} The schematic diagram of the air-liquid EHD problem.}
	\label{fig:geometry}
\end{figure}

Both the liquid and air are considered to be incompressible, Newtonian and perfectly insulating. Following previous theoretical and numerical studies, the governing equations for the flow motion consists of classical continuity and momentum equations:
\begin{equation}
\nabla \cdot \mathbf{u} = 0
\label{eq1}
\end{equation}
\begin{equation}
{{\partial \rho\mathbf{u}}\over{\partial t}}
+\nabla\cdot(\rho\mathbf{u}\mathbf{u})
=\nabla\cdot(\mu\nabla\mathbf{u})
-\nabla p
+\rho\mathbf{g}
+\mathbf{F_e}
+\mathbf{\sigma (\nabla_s\cdot\mathbf{n})}
\label{eq2}
\end{equation}
Here, $\mathbf{u}$ is the velocity of the fluid, $\rho$ is the density, $\mu$ is the dynamic viscosity and $p$ is the pressure. The body force $\rho \mathbf{g}$ refers to the gravity and $\mathbf{\sigma (\nabla_s\cdot\mathbf{n})}$ is the surface tension force where $\sigma$ is the surface tension coefficient, $\nabla_s$ is the surface gradient operator~\cite{melcher1969electrohydrodynamics,castellanos1998nonlinear} and $\mathbf{n}$ is the unit vector normal to the interface and pointing towards the liquid. The electric force $\mathbf{F_e}$ is calculated from the divergence of Maxwell stress tensor in in-compressible fluid\cite{melcher1981continuum}:
\begin{equation}
\mathbf{F_e}
=
\nabla\cdot\left(\varepsilon\mathbf{E} \mathbf{E}-\frac{\varepsilon E^{2}}{2} \mathbb{I}\right)
=
\rho_{e} \mathbf{E}-\frac{1}{2} E^{2} \nabla \varepsilon
\label{eq3}
\end{equation}
where $\rho_e$ is the charge density, $\varepsilon$ is permittivity and $\mathbb{I}$ is the unit tensor. The electric field $\mathbf{E}$ and charge density $\rho_{e}$ in Eq.~(\ref{eq3}) can be obtained by solving the simplified Maxwell equations as given below:
\begin{equation}
\mathbf{E}
=-\nabla \phi
\label{eq4}
\end{equation}
\begin{equation}
\nabla \cdot 
\left( 
\varepsilon \nabla \phi
\right)
= - \rho_e
\label{eq5}
\end{equation}
\begin{equation}
{{\partial \rho_e} \over {\partial t}}
+\nabla \cdot \left(\rho_e \mathbf{u}\right)
+\nabla \cdot \left(\rho_e K \mathbf{E}\right)
-\nabla \cdot \left(D \nabla \rho_e \right)
=0
\label{eq6}
\end{equation}
Here, $\phi$ represented the electric potential, $K$ is the ion mobility and $D$ is the diffusion coefficient. Besides the transient term, the convection, electromigration and diffusion of charge transportation are expressed from left to right in Eq.~(\ref{eq6}). In general, the contribution of charge diffusion is very small when compared to convection and electromigration \cite{castellanos1998electrohydrodynamics}. Thus, a tiny diffusion coefficient ($10^{-30}$) is given to represent the diffusion effect in our study. 

The liquid and air regions are differentiated by solving a scalar transport equation for the local liquid volume fraction $\alpha$ as given below:
\begin{equation}
{{\partial \alpha} \over {\partial t}}+\nabla\cdot \left(\alpha \mathbf{u}\right)
=0
\label{pe}
\end{equation}
The value of $\alpha$ varies from 0 to 1. The regions with $\alpha=1$ and $\alpha=0$ are identified as liquid and air, respectively. This interface model is known as the VOF(Volume Of Fluid) model\cite{VOF} and the physical properties are expressed as a function of local liquid fraction as follows: 
\begin{equation}
P = \alpha P_l + \left(1-\alpha\right) P_a
\label{eq19}
\end{equation}	
where "$P$" represents physical properties including $\rho$, $\mu$, $K$, $\varepsilon$, the subscripts "$l$" and "$a$" indicate the values in the liquid and air regions, respectively. 

The boundary conditions at the electrodes have been shown in Fig.~\ref{fig:geometry}. The left and right sides of the domain are treated as periodic. In addition, the boundary conditions at the interface are as follows\cite{chicon2014stability}:
\begin{widetext}
\begin{equation}
\begin{aligned}
\left [\mathbf{E}\right ] \times \mathbf{n}=0,&\text{ }  [\varepsilon \mathbf{E}] \cdot \mathbf{n}=0,\text{ } [K\mathbf{E}\rho_e] \cdot \mathbf{n}=0, \\
\mu \mathbf{t}& \cdot\left(\nabla \mathbf{u}+\nabla \mathbf{u}^{T}\right) \cdot \mathbf{n}=0, \\
-[p]+\mu \mathbf{n} \cdot\left(\nabla \mathbf{u}+\nabla \mathbf{u}^{T}\right) \cdot \mathbf{n}&+\left[\varepsilon (\mathbf{E} \cdot \mathbf{n})^{2}\right]-\left[\frac{1}{2} \varepsilon E^{2}\right]-\sigma\left(\nabla_{s} \cdot \mathbf{n}\right)=0,
\end{aligned}
\label{eqboundaryInterface}
\end{equation}
\end{widetext}
where$\left [A\right ]$ is the jump from liquid to air for quantity A, and $\mathbf{t}$ is the unit vector tangential to interface.

\subsection{\label{sec:2D}Non-dimensional equations}
The above set of governing equations can be re-written into dimensionless form using the following characteristic scales for length, time, pressure, electric field, charge density and electric current density $D$, $\rho_c {D}^2/\mu_c$, ${\mu_c}^2/(\rho_c {D}^2)$, $V/{D}$, $\varepsilon_c V / {D}^2$ and $\varepsilon_c K_c V^2/{D}^3$, respectively. Here, $\rho_c$, $\mu_c$, $\varepsilon_c$ and $K_c$ are the characteristic physical properties. The obtained non-dimensional system of governing equations\cite{chicon2014stability}: 
\begin{widetext}
\begin{equation}
\nabla \cdot \mathbf{u^*} =0
\label{eq12}
\end{equation}

\begin{eqnarray}
{{\partial \rho^*\mathbf{u^*}}\over{\partial t^*}}
+\nabla\cdot(\rho^*\mathbf{u^*}\mathbf{u^*})
=\nabla\cdot\mu^*\nabla\mathbf{u^*}
-\nabla p^*+U \mathbf{F_e^*}
+\rho^*\mathbf{g^*}
+\frac{\mathbf{g^*}}{Bo} (\nabla_s\cdot\mathbf{n})
\label{eq13}
\end{eqnarray}

\begin{equation}
\mathbf{E^*}
=-\nabla \phi^*
\label{eq14}
\end{equation}
\begin{equation}
\nabla \cdot 
\left( 
\varepsilon^* \nabla \phi^*
\right)
= - \rho_e^*
\label{eq15}
\end{equation}
\begin{equation}
{M\over U^{1/2}}{{\partial \rho_e^*} \over {\partial t^*}}
+{M\over U^{1/2}}\nabla \cdot \left(\rho_e^* \mathbf{u^*}\right)
+\nabla \cdot \left(\rho_e^* K^* \mathbf{E^*}\right)
=0
\label{eq16}
\end{equation}
\end{widetext}
The superscript "*" represents the dimensionless values of the corresponding entities. Adopting the treatment in Ref.~\onlinecite{chicon2014stability}, the properties of liquid ($\rho_l$, $\mu_l$, $\varepsilon_l$, $K_l$) are chosen as characteristic properties ($\rho_c$, $\mu_c$, $\varepsilon_c$, $K_c$). Other non-dimensional parameters that are defined to facilitate the analysis are expressed as follows:
\begin{equation}
{\mathbf{g^*}}=\frac{\rho _{c}^{2} {D}^3 }{\mu_c^2}{\mathbf{g}} ,\text{ } Bo=\frac{\rho_{c}g {D}^2 }{\sigma},\text{ } M=\frac{1}{K_c}\sqrt{\frac{\varepsilon_c }{\rho_c}},\text{ } U=\frac{\varepsilon_c \rho_c V^2}{\mu_c^2}
\label{eq17}
\end{equation}	
Here, $\mathbf{g^*}$ can be treated as a non-dimensional measure for the acceleration due to gravity. $Bo$ is the  Bond number, which represents the ratio of gravitational force to the surface tension force. $M$ is the ratio of the so-named hydrodynamic mobility to ionic mobility. The parameter $U$ represents the ratio of electric force to viscous force and it serves as the driving parameter for the present system. The corresponding non-dimensional boundary conditions are as follows, at electrodes:
\begin{equation}
\begin{aligned}
\mathbf{u^*}=0,\text{ } &  \phi^*=1,\text{ } \rho_e^*=C  \text { at } y^*=1+L^*, \\
\mathbf{u^*}=0,\text{ } &  \phi^*=0 \text { at } y^*=0,
\end{aligned}
\label{eq18}
\end{equation}
at interface:
\begin{widetext}
\begin{equation}
\begin{aligned}
\left [\mathbf{E^*}\right ] \times \mathbf{n}=0,&\text{ }  [\varepsilon^*o \mathbf{E^*}] \cdot \mathbf{n}=0,\text{ } [K^*\mathbf{E^*}\rho_e^*] \cdot \mathbf{n}=0, \\
\mathbf{t}& \cdot\left(\nabla \mathbf{u^*}+\nabla \mathbf{u^*}^{T}\right) \cdot \mathbf{n}=0, \\
-[p^*]+ \mathbf{n} \cdot\left(\nabla \mathbf{u^*}+\nabla \mathbf{u^*}^{T}\right) \cdot \mathbf{n}&+U\left[\varepsilon^* (\mathbf{E^*} \cdot \mathbf{n})^{2}\right]-U\left[\frac{1}{2} \varepsilon^* E^{*2}\right]-\frac{g^*}{Bo}\left(\nabla_{s} \cdot \mathbf{n}\right)=0,
\end{aligned}
\label{eqboundaryInterfaceNondim}
\end{equation}\\
\end{widetext}
where $C=\rho_{e0} D^2/\varepsilon_c V$ is a parameter that serves an indication for the injection strength, $L^*=L/ {D}$ denotes the non-dimensional thicknesses of the air layer. 

\section{\label{sec:3}Numerical methodology}

\begin{figure}
	\centering	
	\includegraphics[width=0.4\textwidth]{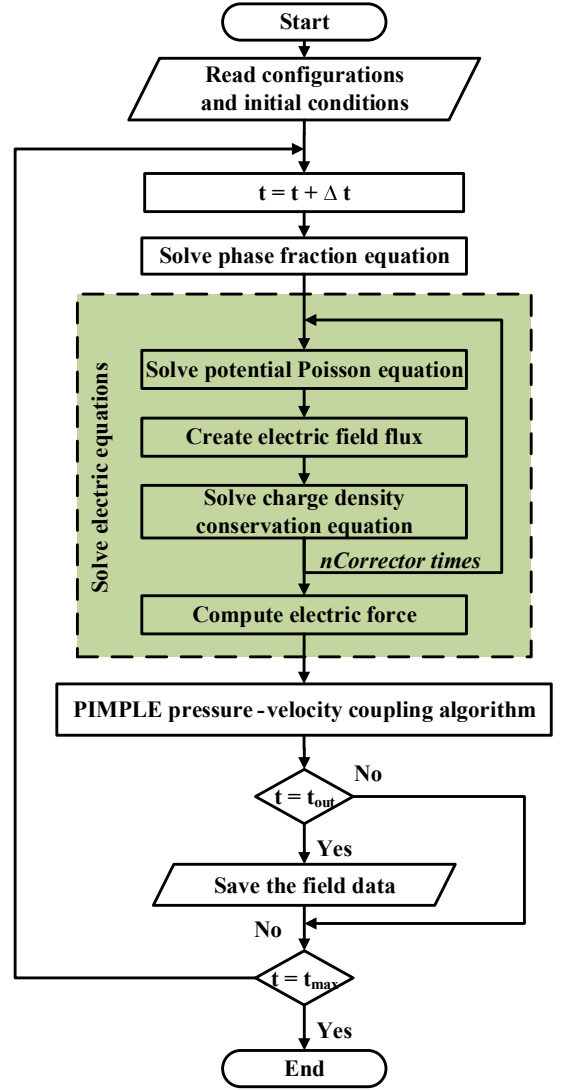}
	\caption{Flow chart of the solution procedure. \textit{nCorrectors times} is the solving times of electric equations.}
	\label{fig:solutionProcedure}
\end{figure}

The numerical model for the two-phase EHD problem presented in this work is built upon the VOF method based finite-volume framework of OpenFOAM\textsuperscript{\textregistered}\cite{Weller1998}. The governing equations for the electric potential, electric field, charge transport and the expression for electric body force term are implemented into the framework. A sequential, iterative solution procedure based on PIMPLE algorithm\cite{PIMPLE} is employed to solve the discrete equations. Since the Poisson's equation and charge density conservation equation are coupled, an iterative sub-loop is designed to ensure full convergence and enhance the solving stability. Fig.\ref{fig:solutionProcedure} presents the overall solution procedure.

The governing equations are discretized using the standard finite-volume procedures available in OpenFOAM\textsuperscript{\textregistered} as described by Moukalled \textit{et al.} \cite{moukalled2016finite}. The time derivatives are discretized using thr Crank-Nicolson scheme with a weighting factor of 0.9. The Laplacian terms present in the governing equations are discretized using a second-order accurate central differencing scheme. A third order cubic scheme is employed to discretize the gradient terms. The convective terms in the momentum and charge density conservation equations are discretized using the third-order QUICK scheme \cite{leonard1979stable} while a second order Total Variation Diminishing (TVD) Van Leer scheme is employed \cite{van1997towards} to discretize the convection term in the phase fraction equation. For the electromigration term, TVD Van Leer scheme is also used to get accurate charge density distribution near the electrode\cite{Selvakumar2021}. OpenFOAM also introduces an interface compression term in Eq.~(\ref{pe}) to sharp the interface\cite{interfaceCom1,interfaceCom2} and this term is discretized with central scheme.

\section{\label{sec:4}Results and Discussion}
This work primarily presents a numerical investigation of EHD stability of a dielectric liquid with horizontal liquid-air interface exposed to a vertical electric field. The process of charge transport under electrohydrostatic equilibrium state and the instability feature as well as the flow pattern are systematically studied. 
\subsection{\label{sec:4A}Choice of parameters and stability analysis}
The geometrical configuration, boundary conditions and governing parameters used in this study are basically adopted from the linear stability of Chicón and Pérez \cite{chicon2014stability}. However, few parameters are set different from Ref.~\onlinecite{chicon2014stability}, in order to reduce the computational expenses. The parameters used in this study are summarized in Table~\ref{tab:table1}.
\begin{table*}[htb]
	\caption{\label{tab:table1}%
		The parameters used in the previous linear stability analysis and the present study
	}
	\begin{ruledtabular}
		\begin{tabular}{ccccccccccc}
			\textrm{} & \textrm{thickness of liquid layer} & \textrm{$K_l/K_a$}  & \textrm{$\varepsilon_l/\varepsilon_a$} & \textrm{$\rho_l/\rho_a$}   & \textrm{$\nu_l$}\footnote{The viscosity in the air layer is not required in the linear stability analysis since it ignores the flow in air. In our simulation, the viscosity of the air is set to be $1.57\times10^{-5}m^{2}/s$} & \textrm{C} & \textrm{Bo}  & \textrm{g*}  & \textrm{M} & \textrm{$L^*$}\\
			\hline
			Linear stability analysis\cite{chicon2014stability} & \multirow{2}{*}{1.2mm} & $5\times10^{-6}$ & \multirow{2}{*}{2.73} & \multirow{2}{*}{800} & \multirow{2}{*}{$5\times10^{-5}m^{2}/s$}   & \multirow{2}{*}{10} & \multirow{2}{*}{0.678} & \multirow{2}{*}{6.78} & \multirow{2}{*}{317.36} & 11.5 \\
			Present study & & $1\times10^{-2}$ &                       &                      &                               &                     &                        &                       &                         & 1.0   
		\end{tabular}
	\end{ruledtabular}
\end{table*}

\begin{figure}[htb]
	\centering	
	\includegraphics[width=0.5\textwidth]{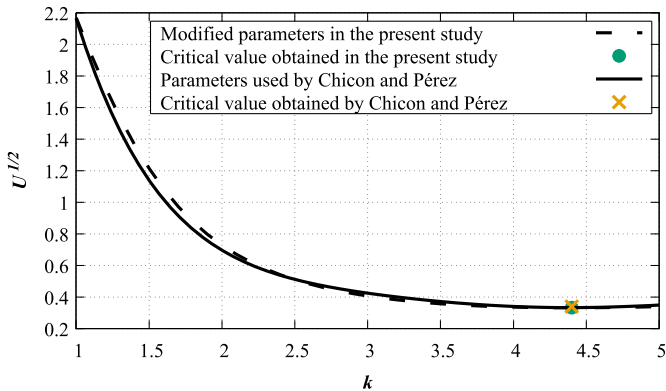}
	\caption{The instability diagrams with the original and changed parameters.}
	\label{fig:instabilityDiagram}
\end{figure}

The first modification is to increase the ratio of ion mobility between the liquid layer and the air layer from $5\times10^{-6}$ used by Chicón and Pérez to $1\times10^{-2}$. The usage of very small ionic mobility ratio leads to a high electromigration flux $K\mathbf{E}$ in the air layer and low flux in the liquid. Thus, a very small time step (around $10^{-4}$ dimensionless time) as well as a long simulation time (around $10^{4}$ to $10^5$ dimensionless time) are required to achieve a divergence free numerical solution. Therefore, a greater ionic mobility ratio is considered to reduce simulation consumption. For the same purpose, the air and liquid layers are set to be of the same thickness in this paper while Chicón and Pérez \cite{chicon2014stability} used an air layer which is 11.5 times thicker than the liquid layer. The instability diagram plotted using the parameters used in Ref.~\onlinecite{chicon2014stability} shows a good agreement with the instability diagram obtained using the modified parameters in this study (Fig.~\ref{fig:instabilityDiagram}). For details of the stability analysis process, please refer to Ref.~\onlinecite{chicon2014stability}. The critical values obtained using the modified parameters are $U_c=0.1106$ and $k_c=4.4$ which are close to the critical values $U_c=0.1155$ and $k_c=4.4$ obtained by using the parameters considered by Ref.~\onlinecite{chicon2014stability}. Thus, it is confirmed that the modified parameters used in this study do not alter the key stability characteristics of the liquid layer. 

To understand why the adjustment of air layer parameters has no significant effect on the stability of the system, the role of air-liquid interface in this problem needs to be discussed. In the stability analysis, the air layer is considered to be in electrohydrodynamic equilibrium because of the large ion mobility in the air layer\cite{chicon2014stability}. Therefore, the instability of the system is dominated by the liquid layer and the interface acts as a flexible electrode plate boundary for the liquid layer. In the mechanics part, the boundary conditions at interface have not changed since the velocity in the air layer is assumed to be zero in the stability analysis. In the electrical part, the change of ion mobility in the air layer will affect the charge accumulation at the interface. However, the charge density in the liquid layer has already become saturated due to the high charge injection intensity. Thus, the limited change of charge accumulation at interface has negligible effect on the electric field characteristics in the liquid layer. As a result, the stability feature shown in Fig.~\ref{fig:instabilityDiagram} has not changed.

\subsection{\label{sec:4B}Electrohydrostatic equilibrium regime}
\begin{table}[htb]
	\caption{\label{tab:table2}%
		The constants in the equations of static solution
	}
	\begin{ruledtabular}
		\begin{tabular}{cccc}
			\textrm{a}  &\textrm{b}\footnote{The values of a, c, d are related to (1-b), which means we have to ensure that the value of b has high accuracy.}   & \textrm{c}  & \textrm{d}                     \\
			\colrule
			0.211738 & $1+1.5038701\times10^{-5}$ & 1.281494 & 0.003663
		\end{tabular}
	\end{ruledtabular}
\end{table}
\begin{figure}[htb]
	\centering
	\includegraphics[width=0.5\textwidth]{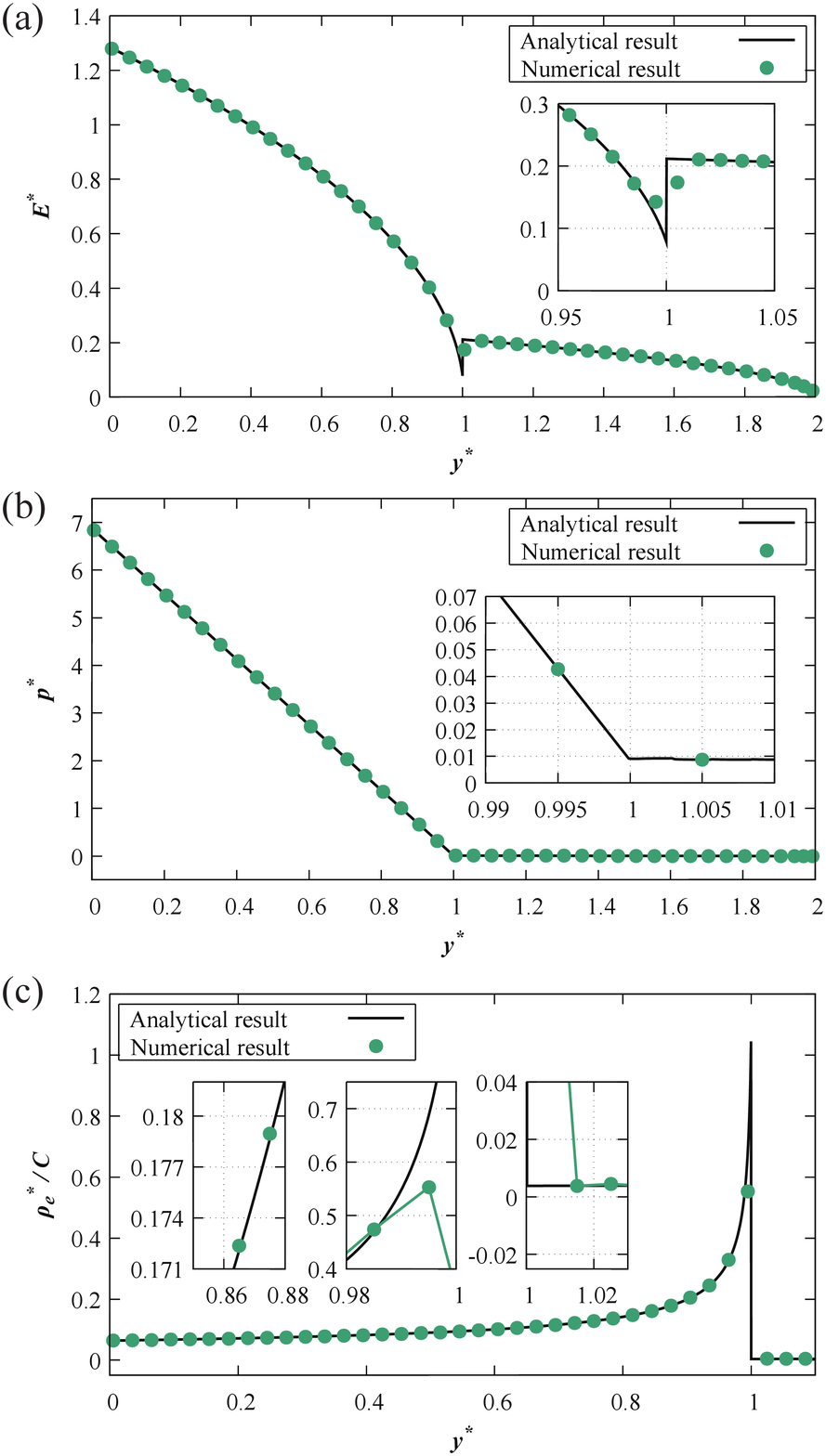}	
	\caption{Comparison of numerical results from present work with analytical solutions in electrohydrodynamic equilibrium condition along y direction. (a) Electric field strength. (b) Pressure. (c) Charge density. }
	\label{fig:staticSolution_a}
	\label{fig:staticSolution_b}
	\label{fig:staticSolution_c}
	\label{fig:staticSolution}
\end{figure}

The present flow problem exhibits an electrohydrostatic regime when the driving parameter $U$ is kept smaller than the critical value. In this regime, the electric body force is weak and cannot induce any motion in the liquid region and thus, the system remains in a rest state. An analytical solution for the charge density, electric field distribution and pressure in the electrohydrostatic regime is given below\cite{chicon2014stability}:

\begin{subequations}
	\begin{equation}
	E_{l}^* =c \sqrt{(1-y^*+d)},\quad E_{a}^* =a \sqrt{(1-y^*+b)} \label{26a}
	\end{equation}
	\begin{equation}
	\rho_{e l}^* =\frac{c}{2 \sqrt{(1-y^*+d)}} ,\quad \rho_{e a}^*=\frac{a\varepsilon_{a}}{2 \varepsilon_{l} \sqrt{(1-y^*+b)}}\label{26b}
	\end{equation}
	\begin{eqnarray}
	p_{l}^*=P_{0 l}^*+\left(g^{*}+U J_{0}\right)(1-y^*) ,\nonumber
	\\
	p_{a}^*=P_{0 a}^*+\left(\rho^{*} g^{*}+U J_{0} \frac{1}{K^{*}}\right)(1-y^*+L)
	\label{eq26c}
	\end{eqnarray}	
	\label{eq26}
\end{subequations}
In Eq.~(\ref{eq26c}), $P_{0 l}^*$ and $P_{0 a}^*$ are the initial parameters determined by the pressure near one of the electrodes and the pressure jump at the interface. $J_0$ is the non-dimensional current density defined as $J_0=K^*\mathbf{E}^*\rho_{e}^*$. $a$, $b$, $c$, $d$ presented in Table~\ref{tab:table2} are constants determined with the parameters listed in Table~\ref{tab:table1}.

\begin{figure}
	\centering	
	\includegraphics[width=0.5\textwidth]{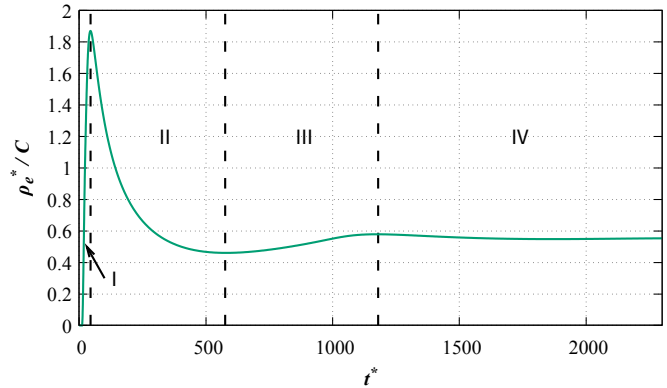}
	\caption{Time evolution of charge density in the mesh cell closest to the interface in the liquid layer.}
	\label{fig:chargeatInterface}
\end{figure}
Fig.~\ref{fig:staticSolution} presents a comparison of our numerical solution with the analytical solution in the electrohydrostatic regime. An nonuniform grid that has been refined near the injection electrode is used in our numerical procedure. There are 200 cells in the vertical direction before refinement which is proved to be mesh independent(see Appendix). The numerical results for the distributions of electric field, pressure and charge density exhibit a good agreement with the analytical solution.

\begin{figure*}[p]
	\centering
	%\subfigbottomskip=0.1pt
	\subfigure{
		\includegraphics[width=0.8\textwidth]{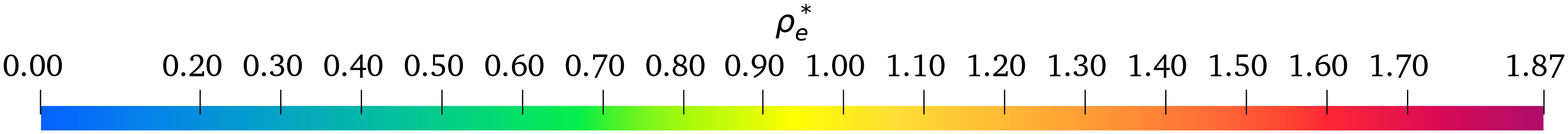}
		
	}
	\setcounter{subfigure}{0}		
	\subfigure[$t^*=30$]{
		\includegraphics[width=0.15\textwidth]{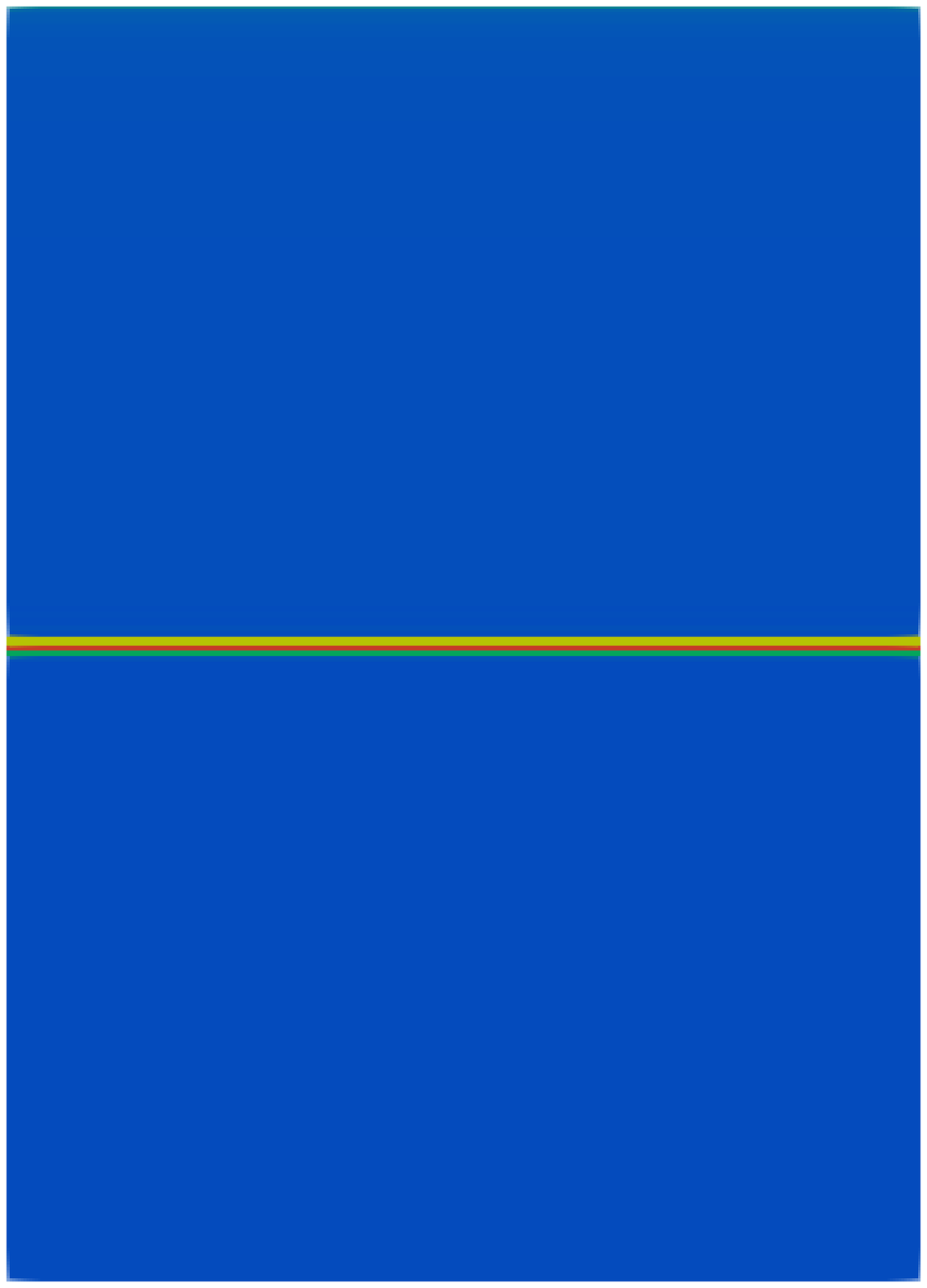}
		\includegraphics[width=0.3\textwidth]{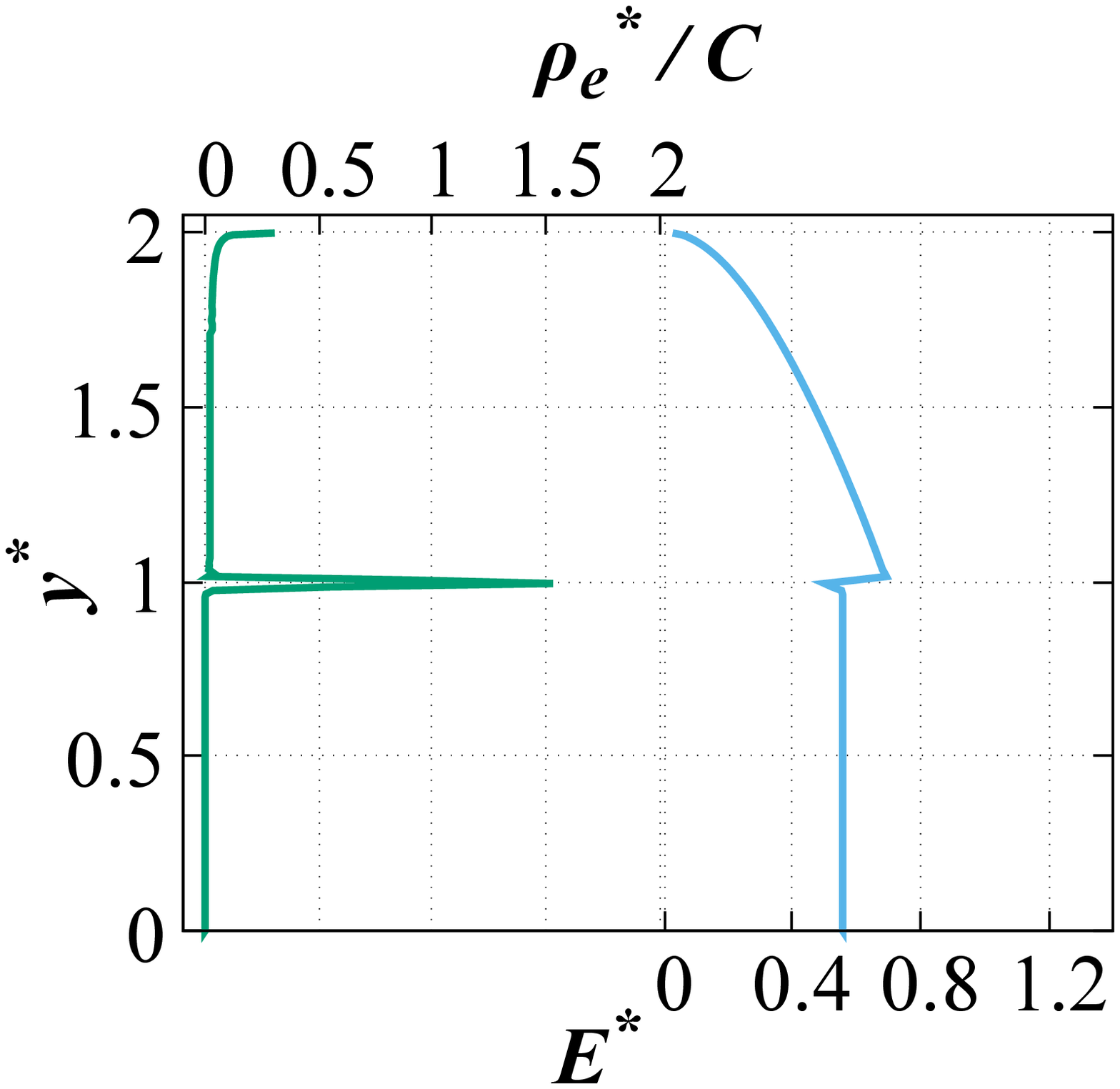}
		\label{fig:chargeInjection_a}
	}
	\subfigure[$t^*=44.4$]{
		\includegraphics[width=0.15\textwidth]{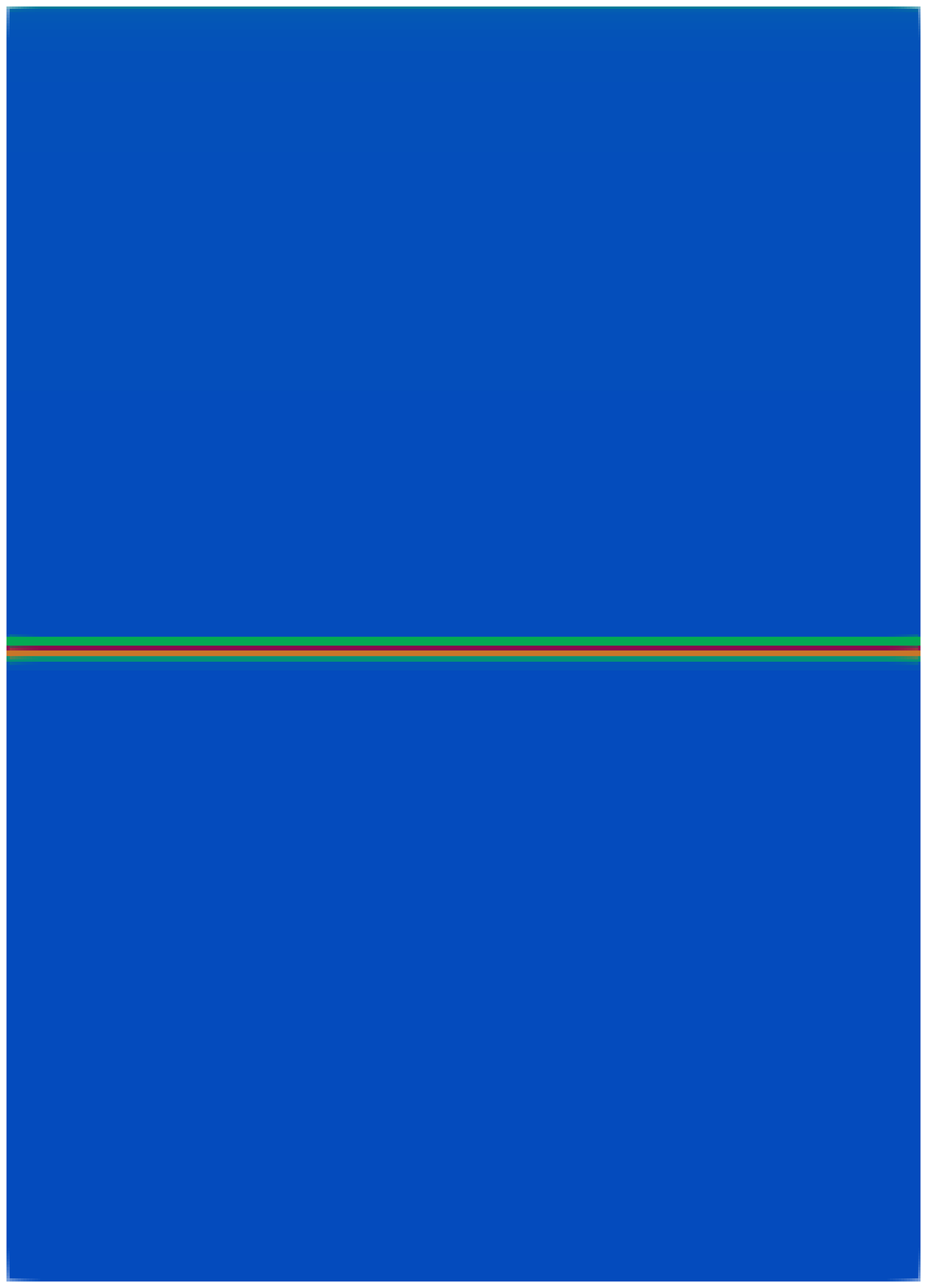}
		\includegraphics[width=0.3\textwidth]{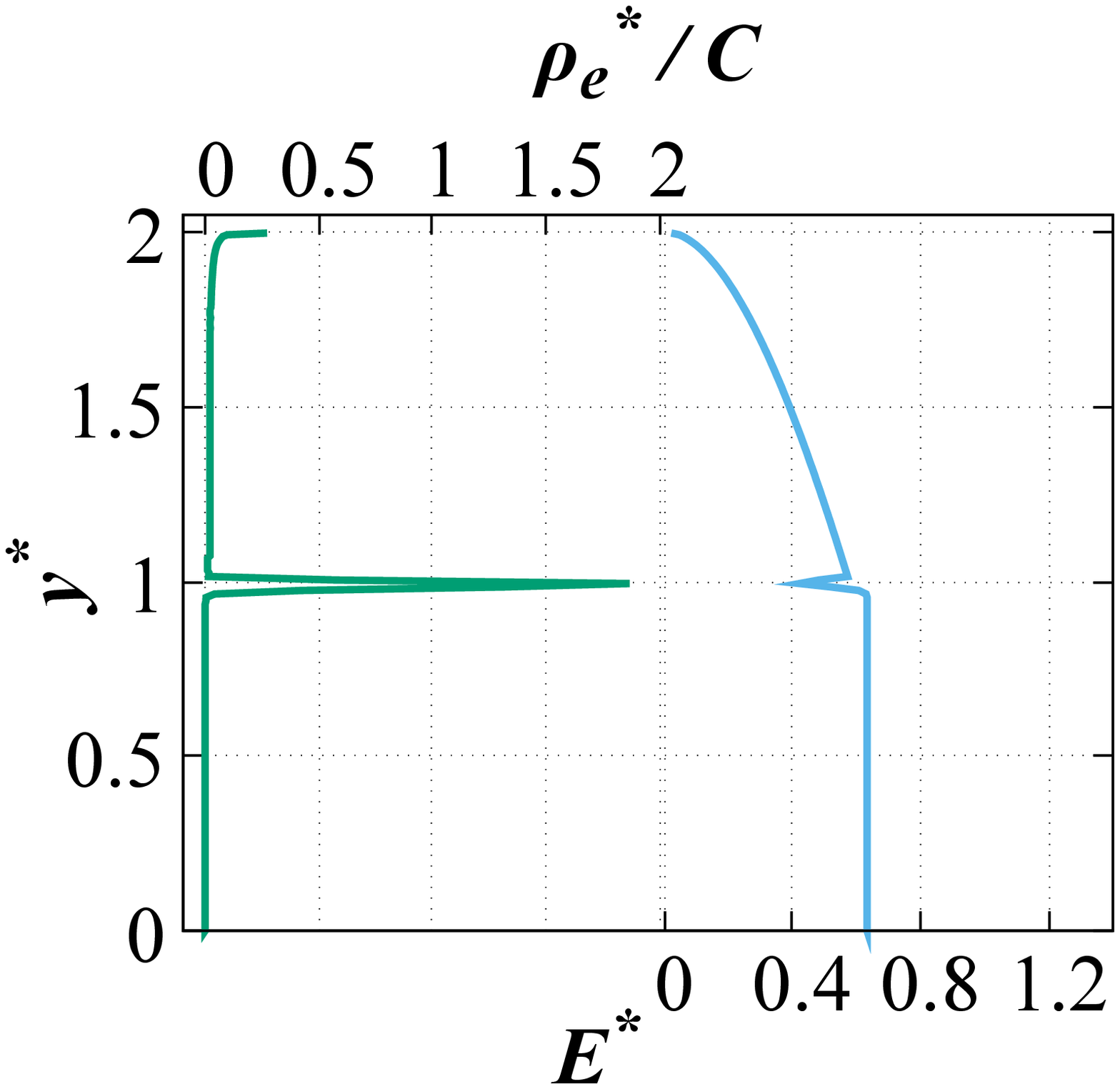}
		\label{fig:chargeInjection_b}
	}
	\\
	\subfigure[$t^*=250$]{
		\includegraphics[width=0.15\textwidth]{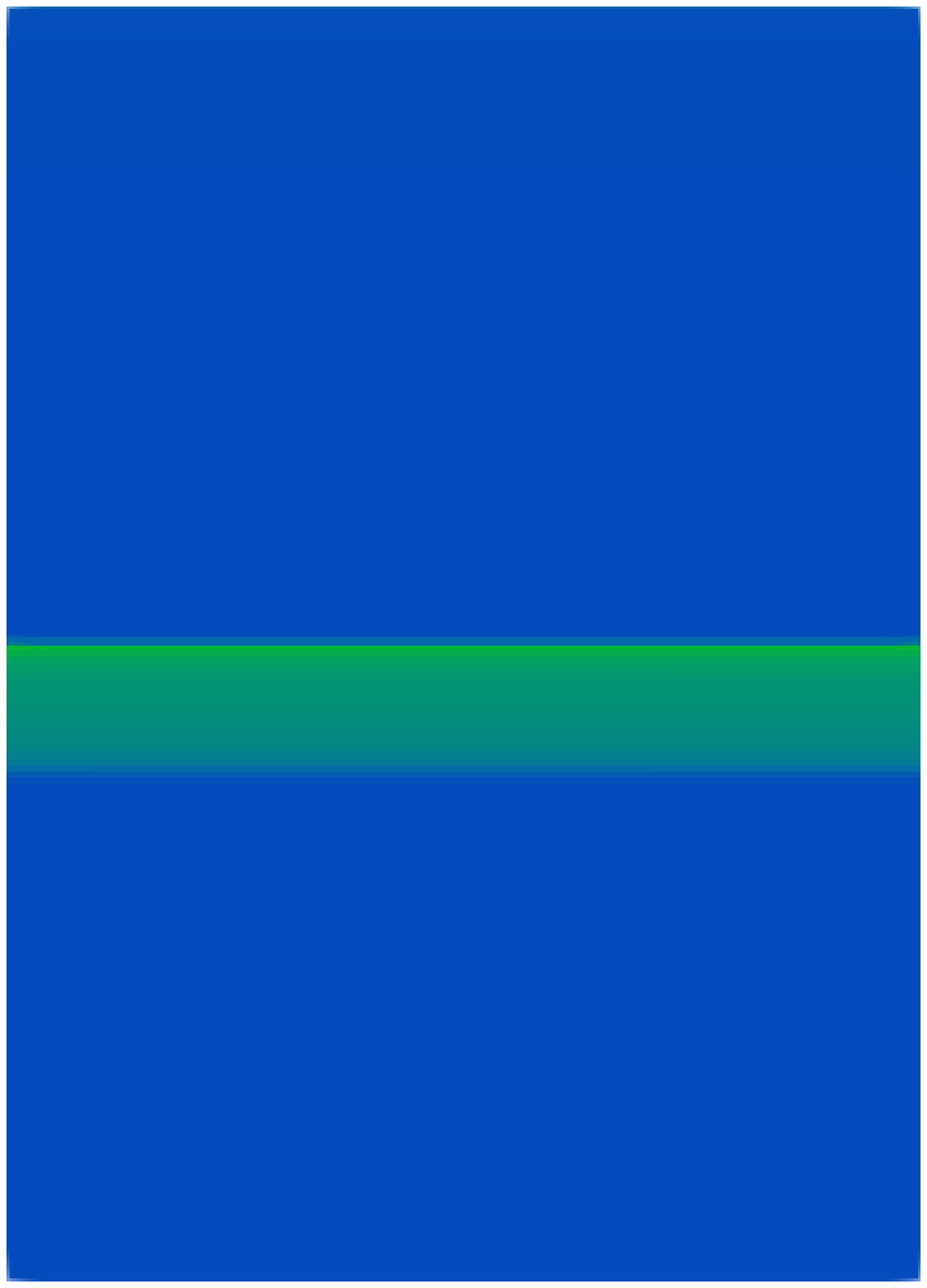}
		\includegraphics[width=0.3\textwidth]{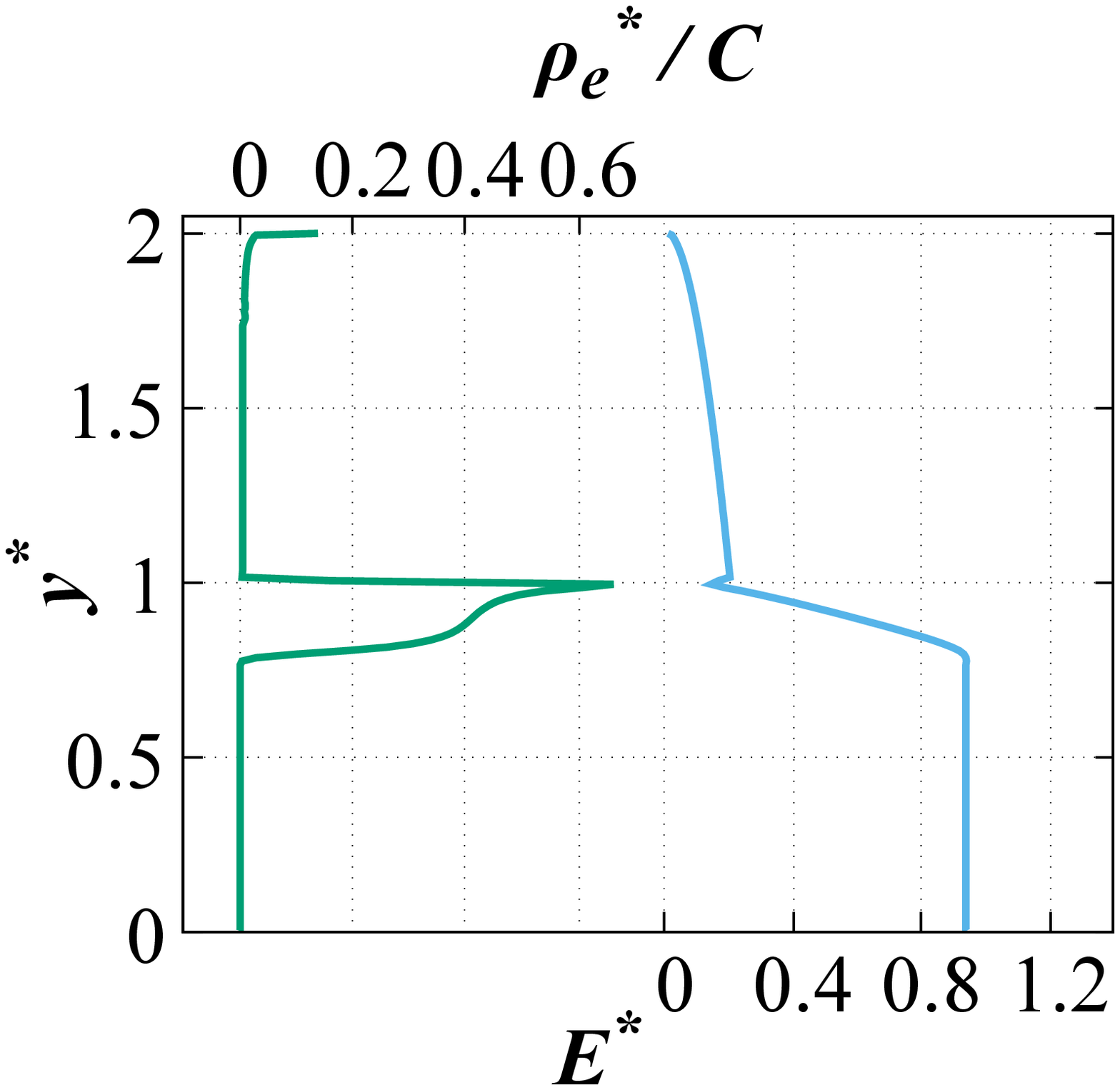}
		\label{fig:chargeInjection_c}
	}
	\subfigure[$t^*=576$]{
		\includegraphics[width=0.15\textwidth]{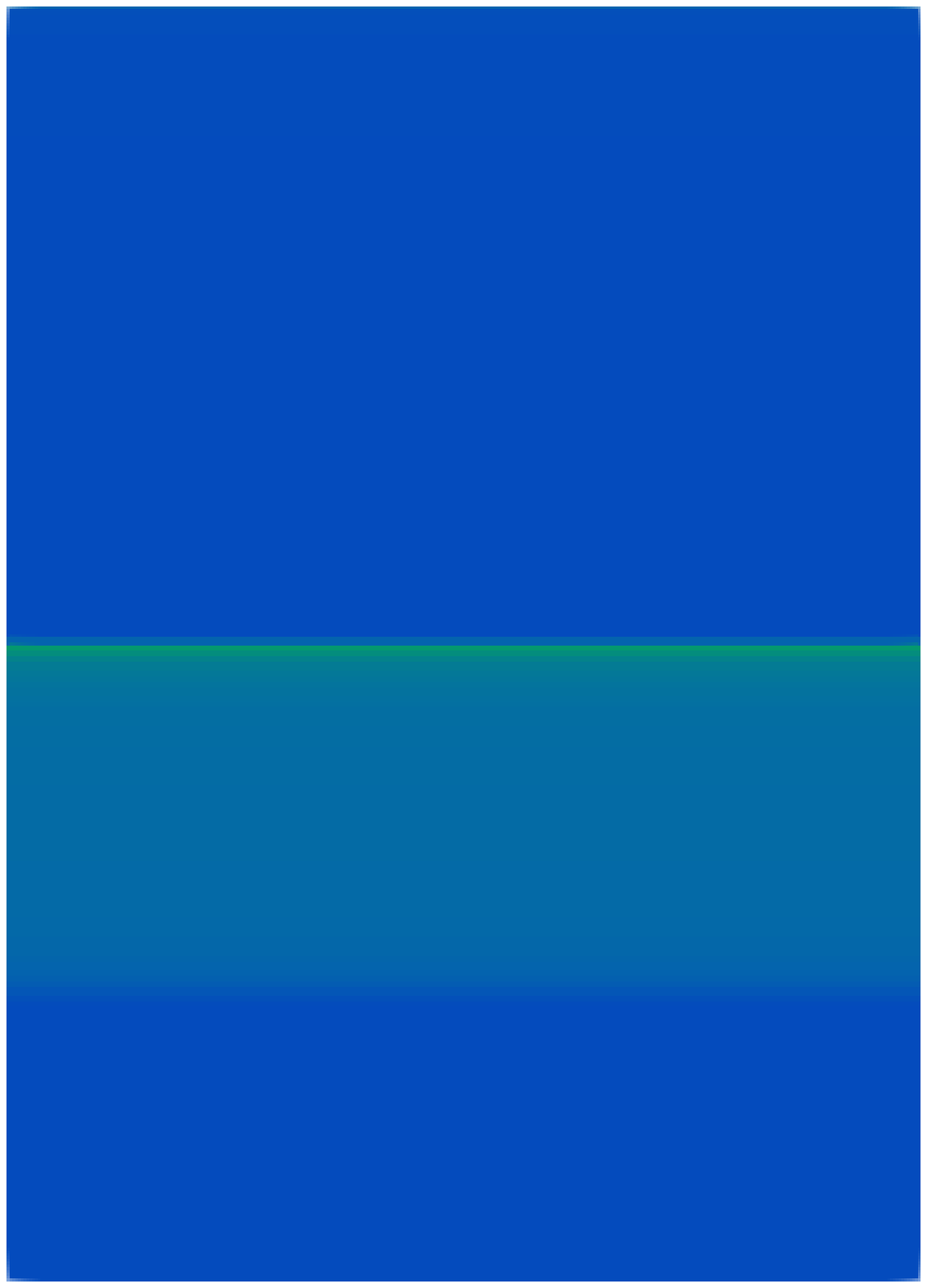}
		\includegraphics[width=0.3\textwidth]{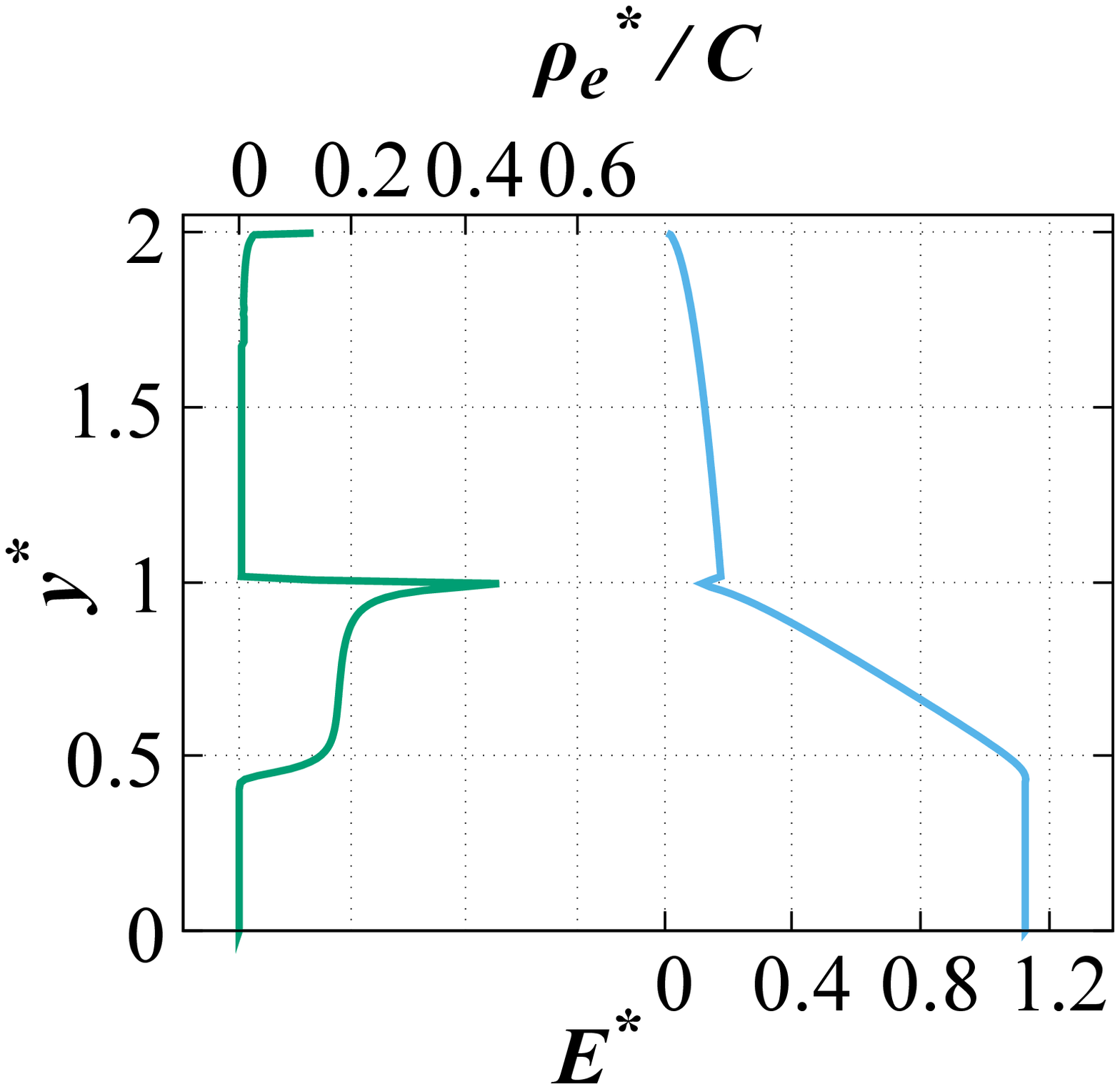}
		\label{fig:chargeInjection_d}
	}
	\\
	\subfigure[$t^*=850$]{
		\includegraphics[width=0.15\textwidth]{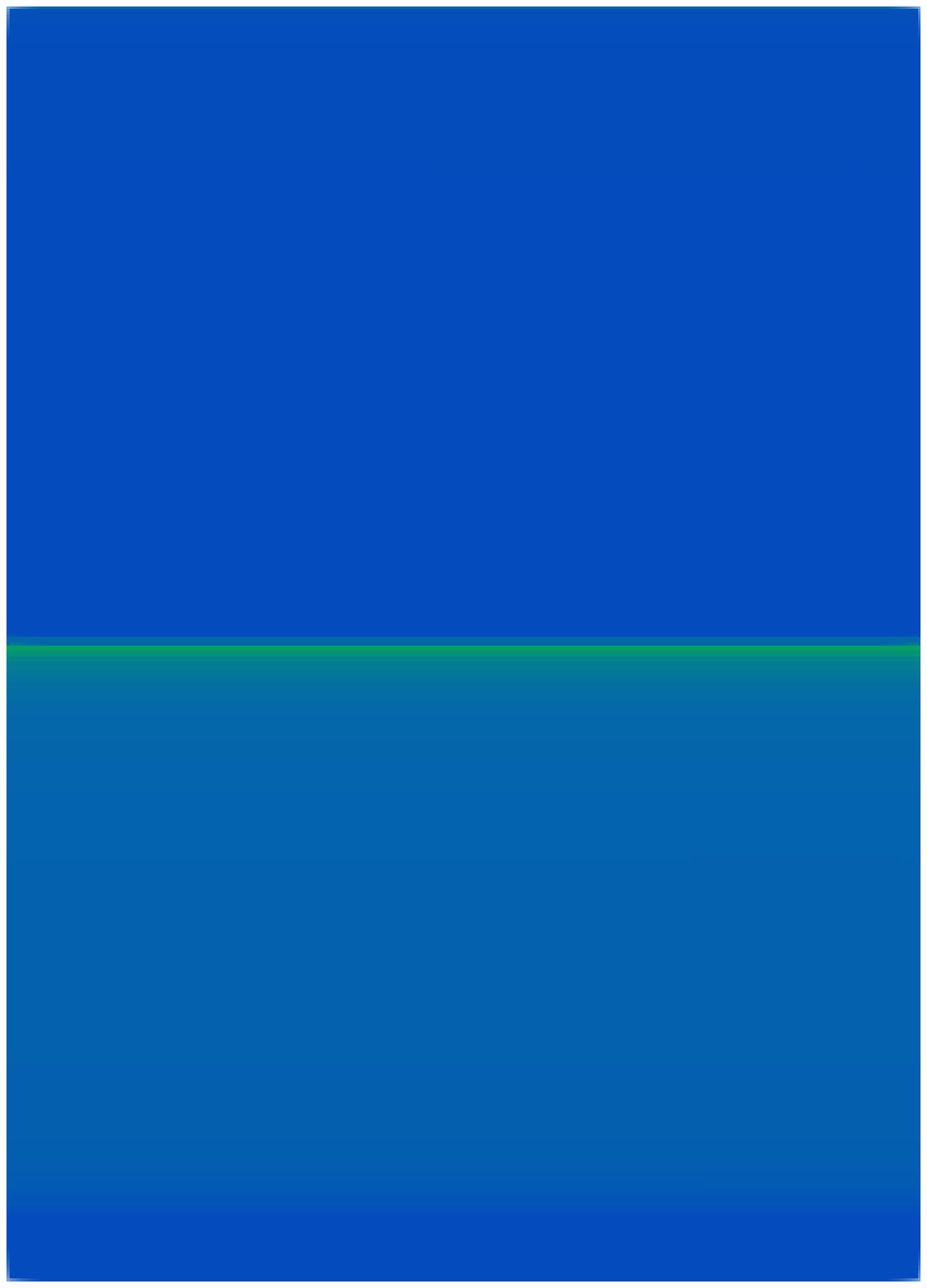}
		\includegraphics[width=0.3\textwidth]{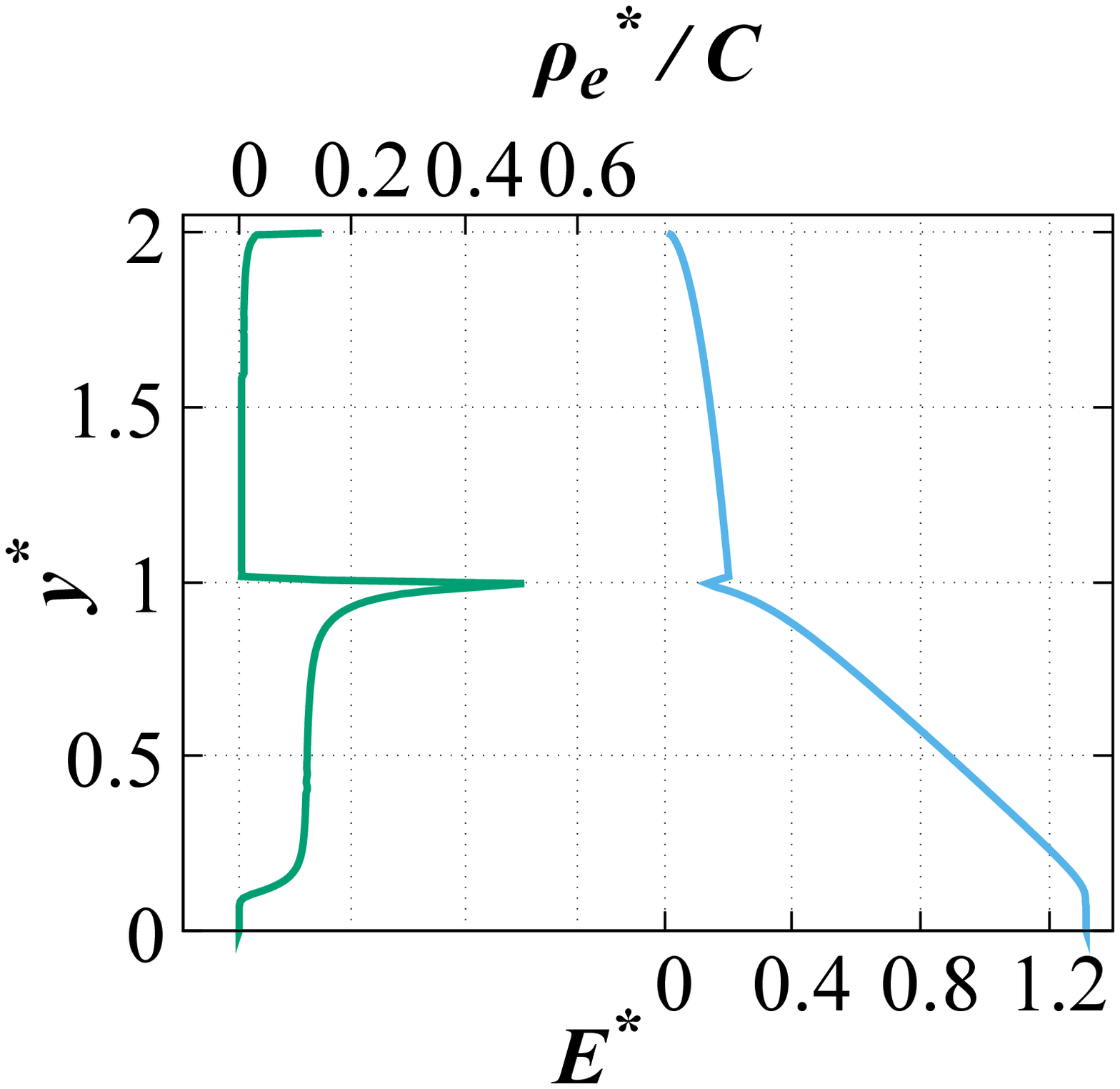}
		\label{fig:chargeInjection_e}
	}
	\subfigure[$t^*=1180$]{
		\includegraphics[width=0.15\textwidth]{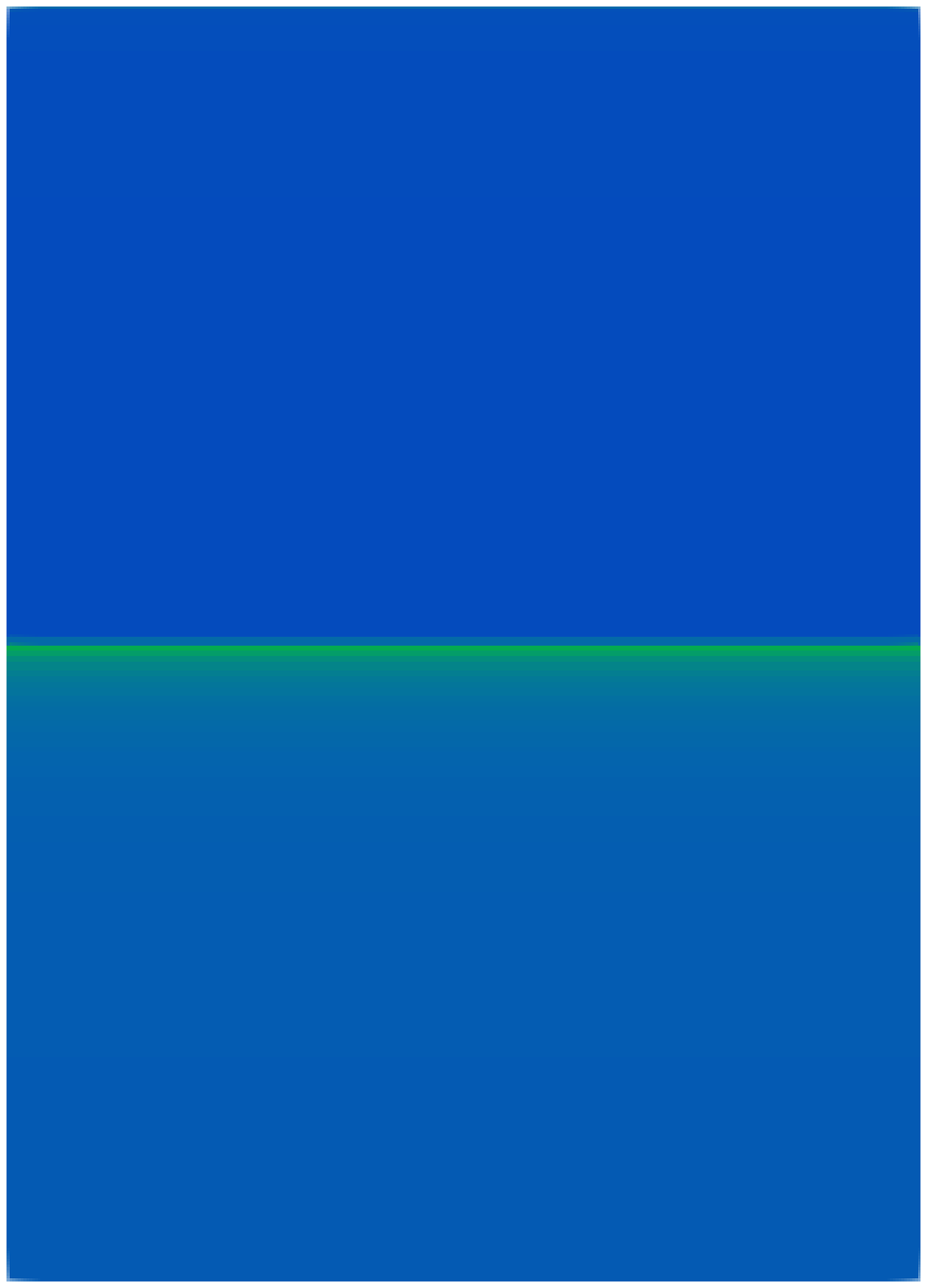}
		\includegraphics[width=0.3\textwidth]{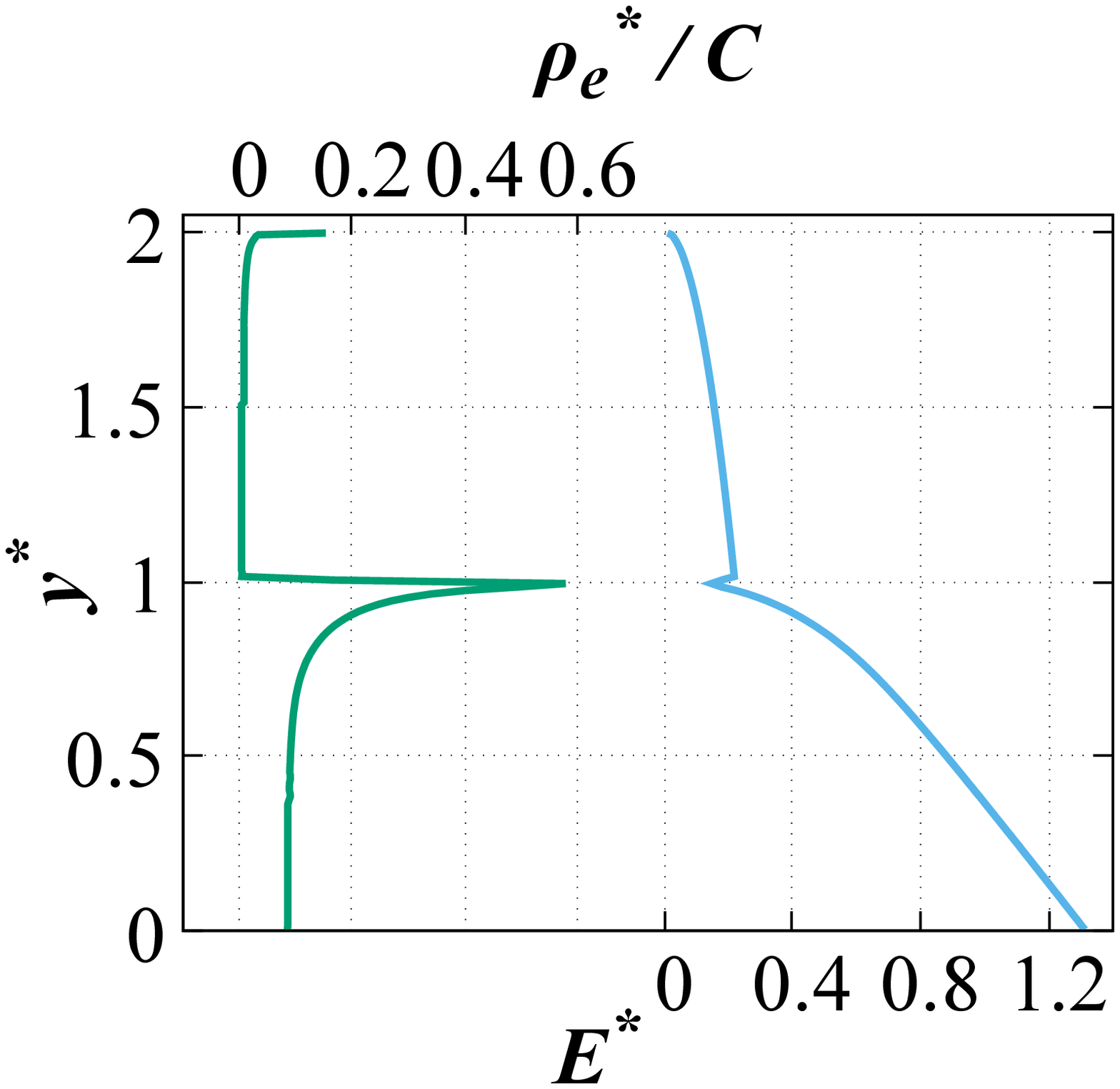}
		\label{fig:chargeInjection_f}
	}
	\\	
	\caption{The charge density(green line) and electric field strength(blue line) distributions during the injection process with U=0.10. When $t^*=44.4$, the charge density near the interface reaches its peak value, while it reaches its valley value at $t^*=576$. }
	\label{fig:chargeInjection}
\end{figure*}

Fig.~\ref{fig:chargeatInterface} presents the charge density evolution with time near the interface, starting with all zero fieds in the bulk. Accordingly, the profiles of $E^*$ and $\rho_e^*$ across the interface at some selected moments are presented in Fig.~\ref{fig:chargeInjection}. In the first stage, for $t^*~<~44.4$ the charge accumulates near the interface and it is observed that the charge does not penetrate much deeper into the liquid layer. For $44.4 \leq t^* \leq 576.0$, the charge accumulation near the interface reaches a very high value and the charge distribution begins to seep into the liquid region. The charge density near the interface depends on the balance between the electromigration velocity in the liquid and air layer. From Eq.~(\ref{eq4}) and Eq.~(\ref{eq5}) we can obtain that once the charge is injected into liquid, the corresponding electric field strength $\mathbf{E}_l$ will increase and finally cause the growth of the migration velocity $K_l\mathbf{E}_l$ in the liquid layer. This increase breaks up the original balance of migration velocities on both sides of the interface and results in the rapid drop of the interface charge density. However, with the further charge transport to the liquid layer, the electric field strength on the liquid side of the interface decreases gradually (as shown in Fig.~\ref{fig:chargeInjection}$(c)$ and Fig.~\ref{fig:chargeInjection}$(d)$). Due to the decreased migration velocity caused by the fall in electric field strength in liquid region, the charge transport velocity in the air layer dominates again. Then, the charge density near the interface again begins to shoot up and this stage is marked as the third region ($576 \leq t^* \leq 1180$, from Fig.~\ref{fig:chargeInjection}$(e)$ to Fig.~\ref{fig:chargeInjection}$(f)$. When the charge reaches the collector electrode, the last stage begins($1180<t^*$). Electric field and charge distribution in liquid layer will undergo an adjusty process to achieve the final state. 

\begin{figure*}[htb]
	\centering
	\includegraphics[width=0.9\textwidth]{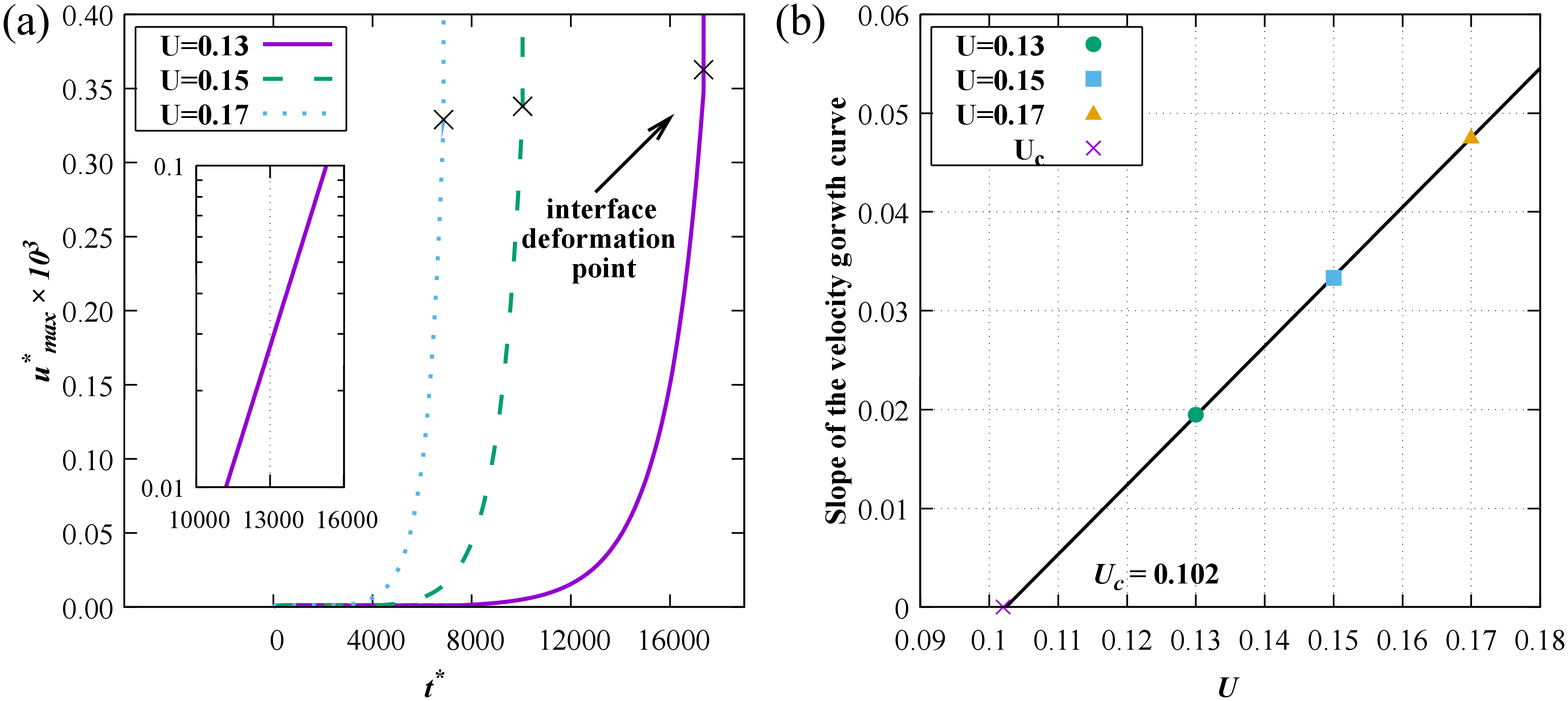}
	\caption{Variation of the maximal velocity $U_{max}$ with different driving parameters $U$. (a) The maximum velocity in liquid layer. Logarithmic coordinates are used to show the exponential growth rate of the velocity in the insert diagram. (b) The linear relationship between the exponential growth rate and driving parameter when $U$ is close to $U_c$. }
	\label{fig:criticalValue_a}
	\label{fig:criticalValue_b}
	\label{fig:criticalValue}
\end{figure*}

\subsection{\label{sec:4C}Stability threshold and flow pattern}
The fluid motion will occur if the driving parameter $U$ is greater than the critical value. To study the flow pattern near the stability criterion, the width of the flow domain in our simulation is considered to be 1.43 times the thickness of the liquid layer, corresponding to the critical wavelength. The variation of the maximum velocity within the liquid layer with respect to the driving parameter $U$ is presented in Fig.~\ref{fig:criticalValue}$(a)$. The liquid layer, which is at an equilibrium state, gradually evolves into a dynamic state. It is observed that the time evolution curve of the maximum velocity undergoes an exponential growth after an initial period of latency. The corresponding linear stability criterion can be estimated using the growth rate of the curve as presented in ref \cite{linearApproach,POF,Pro2}. Following this approach, the velocity growth curve with respect to the parameter $U$ is plotted in Fig.~\ref{fig:criticalValue}$(b)$. The critical value of U is calculated as 0.102, which matches well with the analytical result.

\begin{figure}[htb]
	\centering
	\includegraphics[width=0.5\textwidth]{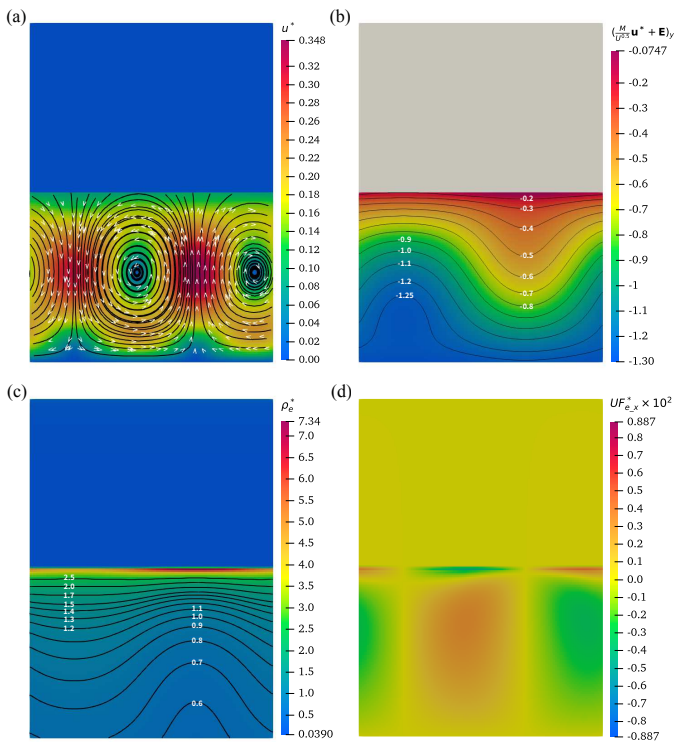}
	\caption{The cloud chart of each dimensionless physical quantity at the moment just before the interface deformation with U=0.13. (a) The distribution of velocity with corresponding streamline. (b) The distribution of total transport velocity with corresponding isoline. Since the values in the air layer are much larger than the value in the liquid layer, the total velocity in air layer is not drawn. (c) The distribution of charge density with corresponding isoline. (d) The distribution of tangential electric force.}
	\label{fig:cloudStandard_a}
	\label{fig:cloudStandard_b}	
	\label{fig:cloudStandard_c}	
	\label{fig:cloudStandard_d}		
	\label{fig:cloudStandard}
\end{figure}

\begin{figure*}
	\centering
	%\subfigbottomskip=0.1pt
	\includegraphics[width=0.75\textwidth]{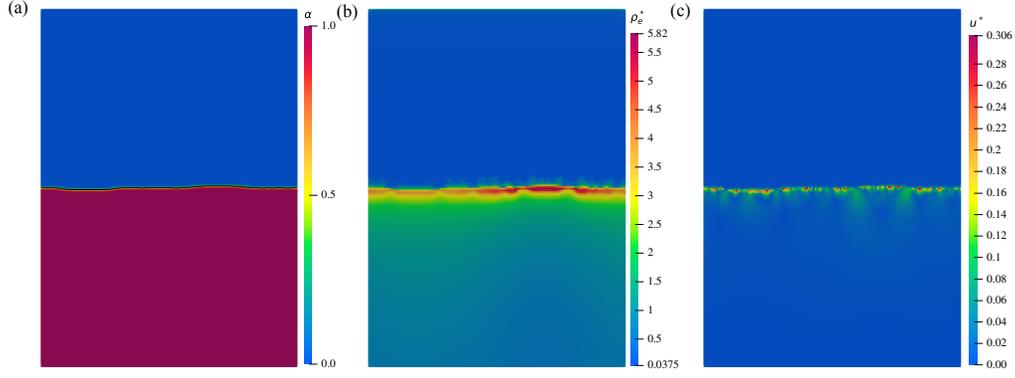}
	\caption{The interface deformation and corresponding charge distribution with U=0.13. (a) The irregular interface deformation. A solid black line is used to distinguish the interface. (b) The distribution of irregular charge. (c) The distribution of velocity. }
	
	\label{fig:afterdeformation_a}
	\label{fig:afterdeformation_b}
	\label{fig:afterdeformation_c}
	\label{fig:afterdeformation}
\end{figure*}

The increase of velocity is essentially related to the intensity growth of the vortex system, as shown in Fig.~\ref{fig:cloudStandard_a}. In the vertical direction, the total ion transportation velocity $K\mathbf{E}+\mathbf{U}$ is enhanced on the side where the vortex flow direction is consistent with the electric field and weakened on the other side (Fig.~\ref{fig:cloudStandard_b}$(b)$). The weakened total migration velocity will give rise to the decrease of charge quantity transported from the interface to the liquid layer, as shown in Fig.~\ref{fig:cloudStandard_c}$(c)$. On the other hand, the amount of charge injected from the air layer to the interface remains unchanged, which finally leads to the increase of charge density near the interface region where the charge injected into liquid is reduced. Therefore, the area with lower charge density in the liquid layer shares the same horizontal position with the zone with higher charge density at the interface. This makes the horizontal component of the electric field that is determined by the gradient of free charge has an opposite direction in liquid layer and interface. Since only Coulomb force $\rho_e\mathbf{E}$ owns non-zero horizontal component when the interface is not deformed, the horizontal electric force reverses its direction near the interface due to the reversed electric field, see Fig.~\ref{fig:cloudStandard_d}$(d)$. The gradient of charge density near the interface is diluted by the lateral flow especially at the junction of the two vortices, which results in a wedge-shaped distribution of electric force near the interface as shown in Fig.~\ref{fig:cloudStandard_d}$(d)$. Since the surface tension is always perpendicular to the interface, the tangential electric force getting larger with the growth of the vortex strength can only be balanced by finite viscous stress, which is a possible mechanism behind interface instability. However, before this mechanism works, the up-flow component in the vortex will deform the interface first. It makes the interface where the highest charge density is seen to protrude upward (see Fig.~\ref{fig:cloudStandard}). The deformation time is marked in the velocity curve in Fig.~\ref{fig:criticalValue_a} and the charge at interface will no longer be stable after the interface's deformation. The unstable charge density leads to more irregular deformation of the interface, as seen in Fig.~\ref{fig:afterdeformation}. After the appearance of the surface deformation, the liquid velocity will jump by several orders, as seen in Fig.~\ref{fig:afterdeformation_c}$(c)$, and the system will become more and more chaotic as time increases. 

\subsection{\label{sec:4D}The charge void region and the effect of the slip boundary at interface}
\begin{figure}
	\centering
	\includegraphics[width=0.5\textwidth]{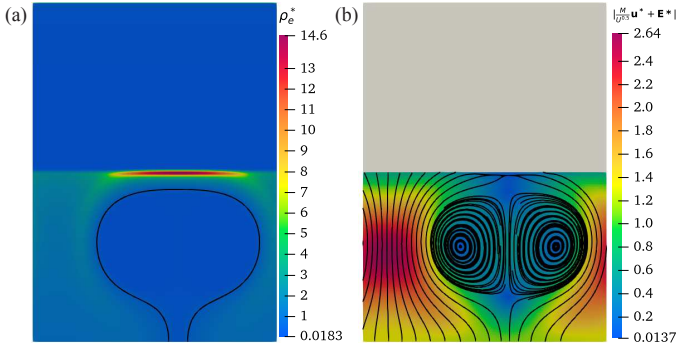}
	%\subfigbottomskip=0.1pt
	%	\subfigure[]{
	%		\label{fig:normalCloud_a}
	%		\includegraphics[width=0.225\textwidth]{figs/normalCloud/charge.eps}
	%	}
	%	\subfigure[]{
	%		\label{fig:normalCloud_b}
	%		\includegraphics[width=0.225\textwidth]{figs/normalCloud/migration.eps}
	%	}
	\caption{The charge and total transport velocity distribution in the domain when U=0.12. (a) The charge void region with the black line as the isoline of $\rho^*=0.5$. (b) The total transport velocity distribution with corresponding streamline.Since the values in the air layer are much larger than the value in the liquid layer, the total velocity in air layer is not drawn. }
	\label{fig:normalCloud_a}
	\label{fig:normalCloud_b}
	\label{fig:normalCloud}
\end{figure}

As we emphasized before, the existence of the interface brings two new factors, deformability and the slip boundary condition, to the system. In order to study the effect of slip boundary conditions at the interface separately, the phase fraction equation Eq.~(\ref{pe}) is blocked in this section. This does not change the boundary conditions at the interface while keeping the interface undeformed. The so-called charge void region shown in the Fig.~\ref{fig:normalCloud_a}(a) then can be captured without the phase update code. This structure is the result of a positive cycle interrupted by interface deformation in early cases and the positive cycle can be described as follows: The weakened charge transport velocity on the side where the vortex flow direction is opposite to the electric field reduces the charge density in the corresponding area. The reduced charge density finally caused the reduction of electric field strength(Eq.~(\ref{eq4}) and Eq.~(\ref{eq5})), which makes the whole transport velocity $K\mathbf{E}+\mathbf{U}$ become smaller again. The final distribution of the whole dimensionless charge transport velocity is given in Fig.~\ref{fig:normalCloud_b} and one can find that the total velocity tends to be zero due to this positive cycle. The self closed streamline of total velocity also shows that no free charge can be entered into and escape from the charge void region. 

\begin{figure*}
	\centering
	%\subfigbottomskip=0.1pt
	%\subfigure[]{
	%	\label{fig:hysteresis_a}
	%	\includegraphics[width=0.4\textwidth]{figs/hysteresis/normal.eps}
	%}
	%\subfigure[]{
	%	\label{fig:hysteresis_b}
	%	\includegraphics[width=0.4\textwidth]{figs/hysteresis/solid.eps}
	%	}
	\includegraphics[width=0.85\textwidth]{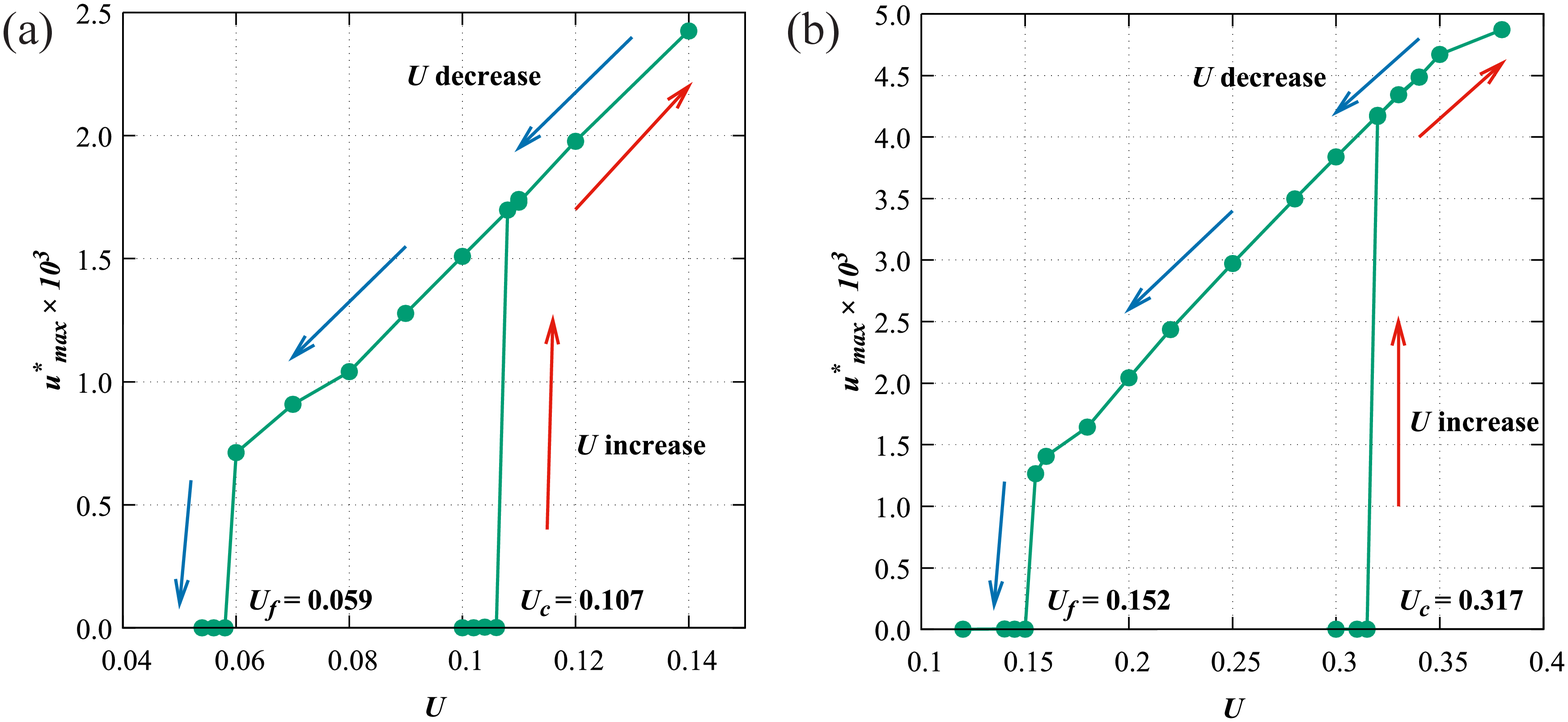}
	\caption{Comparison of hysteresis loop in two-phase and single-phase system. (a) Two-phase system. (b) Single-phase system.}
	\label{fig:hysteresis_a}
	\label{fig:hysteresis_b}
	\label{fig:hysteresis}
\end{figure*}

\begin{figure}
	\centering	
	\includegraphics[width=0.35\textwidth]{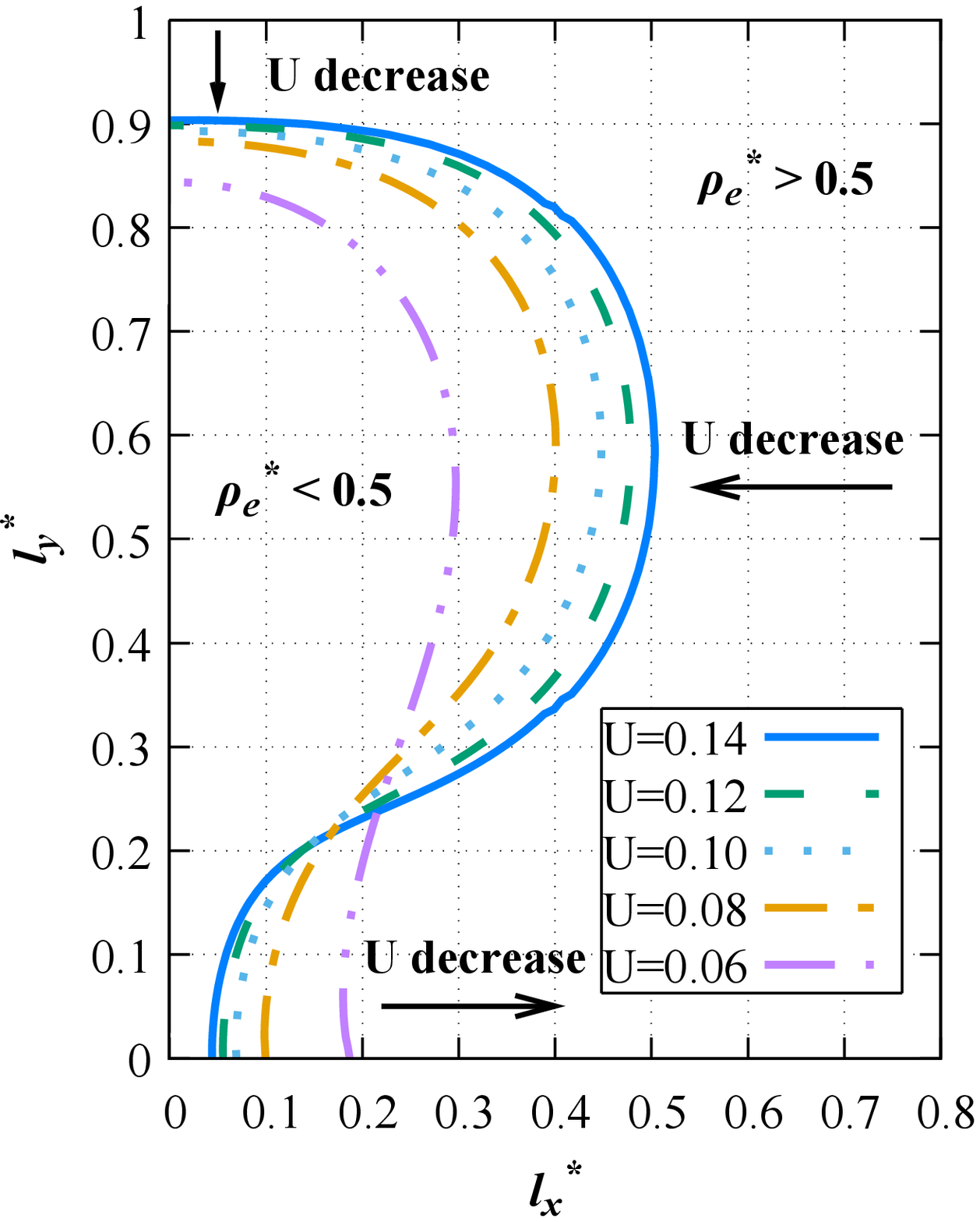}
	\caption{The isoline of $\rho_e^*=0.5$ with different drive parameters.}
	\label{fig:chargeVoidRegion}
\end{figure}

The key characteristic of EHD related convection, the sub-critical bifurcation, is also observed in this study. As shown in Fig.~\ref{fig:hysteresis_a}$(a)$, the system will remain rest when the simulation is started with a small driving parameter $U$. With the increase of $U$, the stable convection cell as well as the charge void region will occur when $U>U_c$. This critical value corresponds to the linear stability criterion. The strength of the flow will gradually weaken when $U$ is decreased from a steady convection until it meets another critical value, where the motion suddenly stops. This criterion is marked as $U_f$ as it is related to finite amplitude instability. We numerically found $U_f=0.059$. Since $U_f$ is smaller than $U_c$, a hysteresis curve is established as presented in Fig.~\ref{fig:hysteresis_a}$(a)$. The hysteresis of the single-phase case is also plotted in Fig.~\ref{fig:hysteresis_b}$(b)$ for comparison. The shape of the charge void region with driving parameters both greater and less than $U_c$ is shown in Fig.~\ref{fig:chargeVoidRegion}. Since the electric torque which drives the flow is proportional to the size of the void region \cite{voidRegionClosetoThecritical}, the void region area owns a non-zero value when $U=0.06$ which is close $U_f$. The width and height of the charge void region become smaller with the decrease of driving parameters while the neck of the region becomes larger. When $U$ lies between the two critical values, the profile changes more sharply than when the driving parameter is less than the $U_c$. This variation is consistent with the results of a single-phase EHD convection \cite{reductionOfChargeVoidRegion}. 

\begin{figure}
	\centering
	%\subfigbottomskip=0.1pt
	%\subfigure[]{
	%	\label{fig:chargeVoidRegionCompare_a}
	%	\includegraphics[width=0.225\textwidth]{figs/chargeVoidRegionCompare/interface.eps}
	%}
	%\subfigure[]{
	%	\label{fig:chargeVoidRegionCompare_b}
	%	\includegraphics[width=0.225\textwidth]{figs/chargeVoidRegionCompare/solid.eps}
	%}
	\includegraphics[width=0.5\textwidth]{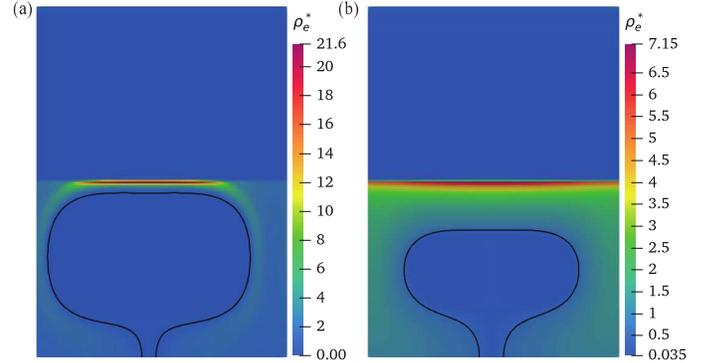}
	\caption{Comparison of charge void region in two-phase and single-phase system with U=0.35. (a) Two-phase system. (b) Single-phase system. }
	\label{fig:chargeVoidRegionCompare_a}
	\label{fig:chargeVoidRegionCompare_b}
	\label{fig:chargeVoidRegionCompare}
\end{figure}

The charge void region in two-phase system exhibits some key differences from a single-phase EHD problem. To illustrate this point, we draw the charge void regions both formed by the single-phase electro-convection and the two-phase flow in Fig.~\ref{fig:chargeVoidRegionCompare}. The two charge void regions are simulated with the same driving parameters. Key differences between two charge void regions can be summarized as follows. Firstly, the void region in the two-phase system has a larger size, which means the convection strength is greater. This can also be related to Fig.\ref{fig:hysteresis} where the single-phase case shows a much higher linear and nonlinear critical values. Second, the upper edge of the charge void region coincides with the interface in the two-phase system while there exists a notable gap between the void region and the interface(electrode) in a single-phase case. This can be explained by the velocity boundary conditions at interface and rigid electrode. The interface acts as a slip boundary condition since the velocity is continuous between the movable air and liquid, which allows vortex structure in the liquid layer extend to the interface. While the viscous effect near the no-slip electrode forces the vortex to stay away from the injection electrode for a single-phase problem. This viscous effect is also the main reason for the smaller charge void region in single-phase flow. Third, the charge density at the interface exhibits a much higher peak value and horizontal gradient in air-liquid system. This phenomenon can be understood from two aspects. On the one hand, the convection strength is much larger in the presence of an interface, which makes the flow have a stronger influence on charge distribution. On the other hand, since the charge can't be transported into charge void region due to the self-closed total migration velocity (Fig.~\ref{fig:normalCloud_b}$(b)$), the interface closely connected with the void region in air-liquid system cannot inject enough charge to the liquid layer, which makes it easier for that part of interface to accumulate more charge from the upper layer. Overall, it can be inferred that the absence of viscous effect at interface allows for a finite velocity there, which makes it easier for the flow to develop and generate a larger flow intensity, resulting in a smaller critical value than single-phase system. The structure of the charge void region is also effected by the stronger flow and the velocity at interface.

\section{\label{sec:6}Concluding remarks}
In this paper, two-dimensional numerical simulation was performed to study the electrohydrodynamic (EHD) instability with a planar layer of air-dielectric liquid subjected to unipolar injection. This fundamental two-phase EHD problem has been investigated with the linear stability tool, and this study presents direct numerical results for the first time. A finite volume solver with the electrostatic equations implemented in the open-source platform OpenFOAM® was developed, and the volume of fluid (VOF) model was used to deal with the liquid-air interface. The solver was first validated by the electrohydrostatic equilibrium solution. Then numerical results with the onset of flow motion and flow structure were presented. The numerically obtained critical value for the linear stability matches well with the theoretical values. Once the flow develops, the interface deformation becomes irregular, and the resulting flow structure and charge void region is chaotic. To understand the effects of slip boundary at interface, the stable convection was obtained with blocked interface. In this case, the finite amplitude criterion, which corresponds to the stop of flow motion, was obtained as Uf = 0.059. A hysteresis loop linking the linear and finite amplitude criteria was also determined in the bifurcation diagram. The distribution of tangential electric force near the interface and its formation mechanism are described. It is considered that this distribution may have a significant contribution to the interface deformation. The lack of viscous effect at the free interface accounts for the smaller instability threshold in two-phase system compared with the single-phase case. The presence of interface also enlarges the size of charge void region and shifts its position closer to the interface. 

Though the present study and also previous stability analysis\cite{chicon2014stability} are inspired by the rose-window instability, the experimental condition of needle-plate electrode configuration and much higher values of driving parameter bring some inherent difference and challeng. The 3D simulation of rose-window instability phenomenon can be a future work.

\begin{acknowledgments}
	Jian Wu acknowledges financial support by the National Natural Science Foundation of China via Grant No.11802079 and 51906051.	Alberto T. Pérez acknowledges financial support by the Spanish Ministerio de Ciencia, Innovación y Universidades under Research Project No. PGC2018-099217-B-I00, by the Ministerio de Economía y Competitividad under Research Project No. CTQ2017-83602-C2-2-R, and Junta de Andalucía under research project 2019/FQM-253.
\end{acknowledgments}

\appendix

\section{\label{sec:app} Grid independent study}
\begin{figure}
	\centering	
	\includegraphics[width=0.5\textwidth]{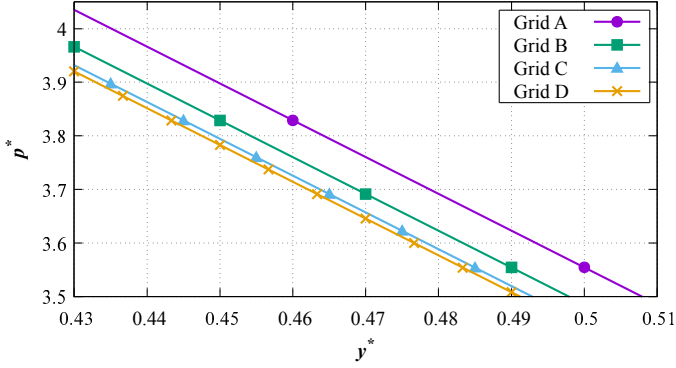}
	\caption{The pressure distribution in liquid layer with different grids.}
	\label{fig:meshIndependent}
\end{figure}
\begin{table*}[]
	\caption{\label{tab:table3}
		The numerical result with different grids\footnote{The relative error is defined as $[ (J^*,p^*,\int_{liquid} p^*dy^*)_{c}-(J^*,p^*,\int_{liquid} p^*dy^*)_{r})]/(J^*,p^*,\int_{liquid} p^*dy^*)_{r}$ where the subscript "c" is the value of the current grid and "r" is the value of the rougher grid.}
	}
	\begin{ruledtabular}
		\begin{tabular}{ccccc}
			& grid A      & grid B      & grid C      & grid D      \\
			\hline
			$J^*$                         & 0.81696 & 0.81918 & 0.82017 & 0.82039 \\
			relative error of $J^*$       & -       & 0.27\%  & 0.12\%  & 0.03\%  \\
			relative error of $p^*$       & -       & 1.96\%  & 1.01\%  & 0.32\%  \\
			$u^*_{max}\times 10^3$        & 3.08748 & 5.48995 & 6.77753 & 6.62854 \\
			relative error of $u^*_{max}$ & -       & 77.81\% & 23.45\% & 2.20\%       
		\end{tabular}
	\end{ruledtabular}
\end{table*}

Four different orthogonal grids named as (1) grid A - (36$\times$50), (2) grid B - (72$\times$100), (3) grid C - (144$\times$200), (4) grid D - (216$\times$300) with mesh refinement at $L/15$ away from the injection electrode are used to perform the simulations. Dimensionless current density $J^*=K^*E^*\rho_e^*$, dimensionless pressure distribution under static solution with the drive parameter U=0.10 as well as the maximum velocity in liquid layer with U=0.35 (the case of Fig.~\ref{fig:chargeVoidRegionCompare}) are used as electricity and mechanics evaluation indexes. The results are shown in Fig.~\ref{fig:meshIndependent} and Table~\ref{tab:table3}.

With the increase of grid density, the current density and pressure show little changes while the maximum velocity improves a lot. When the grid is encrypted from grid C to grid D, the result improvement is very limited. Considering both the accuracy and the consumption of simulation resources, grid C is chosen as the final grid for simulations.

%\nocite{*}
\bibliographystyle{unsrt}
\bibliography{myPRF}% Produces the bibliography via BibTeX.

\end{document}